\documentclass[12pt,english,english]{article}
\usepackage[T1]{fontenc}
\usepackage{babel}
\usepackage{graphics}

\makeatletter

\providecommand{\LyX}{L\kern-.1667em\lower.25em\hbox{Y}\kern-.125emX\@}

 \newcommand{\lyxaddress}[1]{
   \par {\raggedright #1 
   \vspace{1.4em}
   \noindent\par}
 }

\usepackage{bookman}
\usepackage[T1]{fontenc}
\usepackage[latin1]{inputenc}
\usepackage{amsmath}
\usepackage{graphicx}

\makeatletter

\def\fnum@table{\tablename~{\bf\thetable}}
\def\fnum@figure{\figurename~{\bf\thefigure}}
\def\tablename{\footnotesize{\bf Table}}
\def\figurename{\footnotesize{\bf Figure}}

\AtBeginDocument{
  
}

\usepackage{babel}
\makeatother

\makeatother
\begin{document}

\title{\textbf{Hadron Production in Proton-Proton Scattering in NE}\textbf{\huge X}\textbf{US
3}}

\author{F.M. Liu\( ^{1,2,4} \)%
\thanks{Fellow of the Alexander von Humboldt Foundation
}, T. Pierog\( ^{3,4} \), J. Aichelin\( ^{4} \), K. Werner\( ^{4} \) }

\maketitle

\lyxaddress{\noindent \( ^{1} \) \emph{\small Institute of Particle Physics,
Central China Normal University, Wuhan, China.} \\
 \emph{\small \( ^{2} \) Institut fuer Theoretische Physik, JWG Frankfurt
Universitaet, Frankfurt, Germany} \\
 \textit{\small \( ^{3} \) Forschungszentrum Karlsruhe, Institut
fuer Kernphysik, Karlsruhe, Germany}\\
 \emph{\small \( ^{4} \) Laboratoire SUBATECH, University of Nantes
- IN2P3/CNRS - Ecole des Mines de Nantes, Nantes, France}}

Using the recently introduced model NE{\large X}US 3, we calculate
for \( pp, np \) and \( \bar{p}p \) collisions the excitation function
of particle yields and of average transverse momenta of different
particle species as well as rapidity, \( x_{F} \) and transverse
momentum distributions. Our results are compared with available data 
in between \( \sqrt{s} \) = 5 GeV and 65 GeV.
We find for all observables quite nice agreement with data what make
this model to a useful tool to study particle production in elementary
hadronic reactions.

\section{Introduction}

The recently published \( \Omega  \) and \( \bar{\Omega } \) data
in pp collisions at 158 GeV by the NA49 collaboration have invalidated
all customary quark-diquark string models \cite{nexn,omeg}, which have been
employed since long time to describe pp as well as AA collisions.
In all those models, the valence quarks are taken as the end points
of the strings, which provokes that due to the string topology \cite{omeg}
more \( \bar{\Omega } \) than
\( \Omega  \) are produced, in disagreement with the data. This effect
has been verified in detailed calculations.

In the new NE{\large X}US 3 model, the observed particles are produced by
two sources: a) strings which are formed by sea (anti)quarks and which
are therefore symmetric with respect to the exchange of a particle
and an antiparticle and b) excited remnants, which decay statistically.
Whereas the string produces an equal number of \( \Omega  \) and
\( \bar{\Omega } \), the remnant favors due to its finite baryon number 
the production of \( \Omega  \),
in accordance with the data.

In this paper, we calculate the results of the NE{\large X}US 3 model
for particle yields, rapidity distributions, transverse momentum distributions
and average transverse momenta. Where experimental data are available
we compare with them. We will show that the prediction of this model,
once the necessary parameters are fixed to describe the CERN SPS pp
data at 158 GeV and the excitation function of the multiplicity of
charged particles, are in good agreement with the experimental data
down to beam energies as low as 10 GeV . A similar, however less complete,
comparison between data and string fragmentation models  
has recently be performed by H. Weber et al.
\cite{weber}. They compare the URQMD and HSD model with data of nonstrange 
or single strange particles.

\section{NE{\huge X}US 3}

NE{\large X}US 3 is a self-consistent multiple scattering approach to
proton-proton and nucleus-nucleus scattering at high energies. The
basic feature is the fact that several elementary interactions, referred
to as Pomerons, may happen in parallel. We use the language of Gribov-Regge
theory to calculate probabilities of collision configurations (characterized
by the number of Pomerons involved, and their energy) and the language
of strings to treat particle production. 
\begin{figure}
{\centering \includegraphics[  scale=0.7]{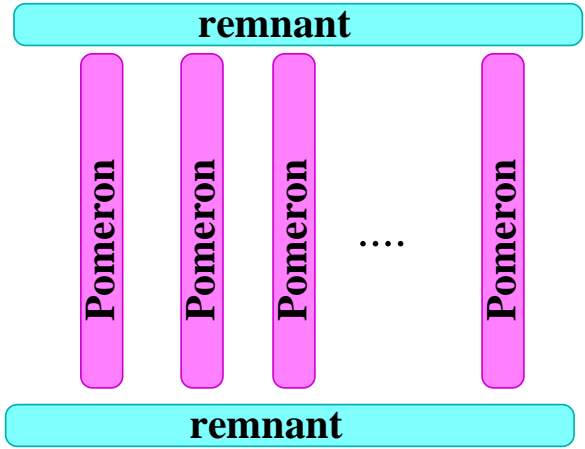}\par}

\caption{\label{mult} {\small Multiple elementary interactions (Pomerons)
in} NE{\large X}US {\small .}}
\end{figure}

We treat both aspects, probability calculations and particle production,
in a consistent fashion: \emph{In both cases energy sharing is considered
in a rigorous way} \cite{nexo}\emph{, and in both cases all Pomerons
are identical.} This is one new feature of our approach. Another new
aspect is the necessity to introduce remnants: The spectators of each
baryon form a remnant, see Fig. \ref{mult}. They will play an important
role on particle production in the fragmentation region and at low
energies (\( E_{Lab}= \)40-200 GeV). In the following we discuss
some more details of our approach.

\begin{figure}
{\centering \includegraphics[  scale=0.4]{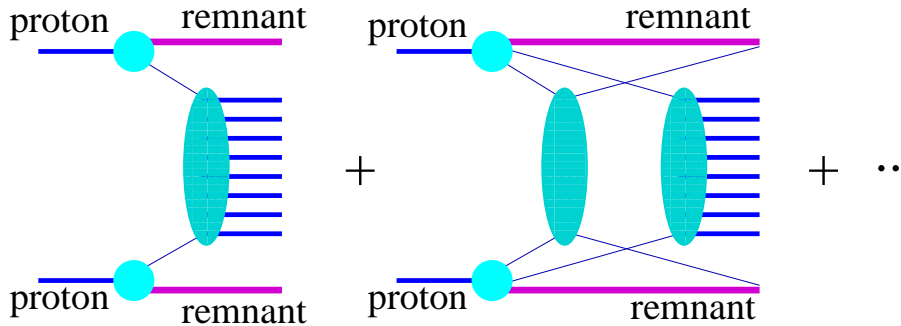}\par}

\caption{\label{t7}Inelastic scattering in pp. Partons from the projectile
or the target proton interact via elementary interactions (the corresponding
produced particles being represented by horizontal lines), leaving
behind two remnants. }
\end{figure}

We first consider inelastic proton-proton scattering. We imagine an
arbitrary number of elementary interactions to happen in parallel,
where an interaction may be elastic or inelastic, see Fig. \ref{t7}.
The inelastic amplitude is the sum of all such contributions in with
at least one inelastic elementary interaction is involved. 

\begin{figure}
{\centering (a)\includegraphics[  scale=0.4]{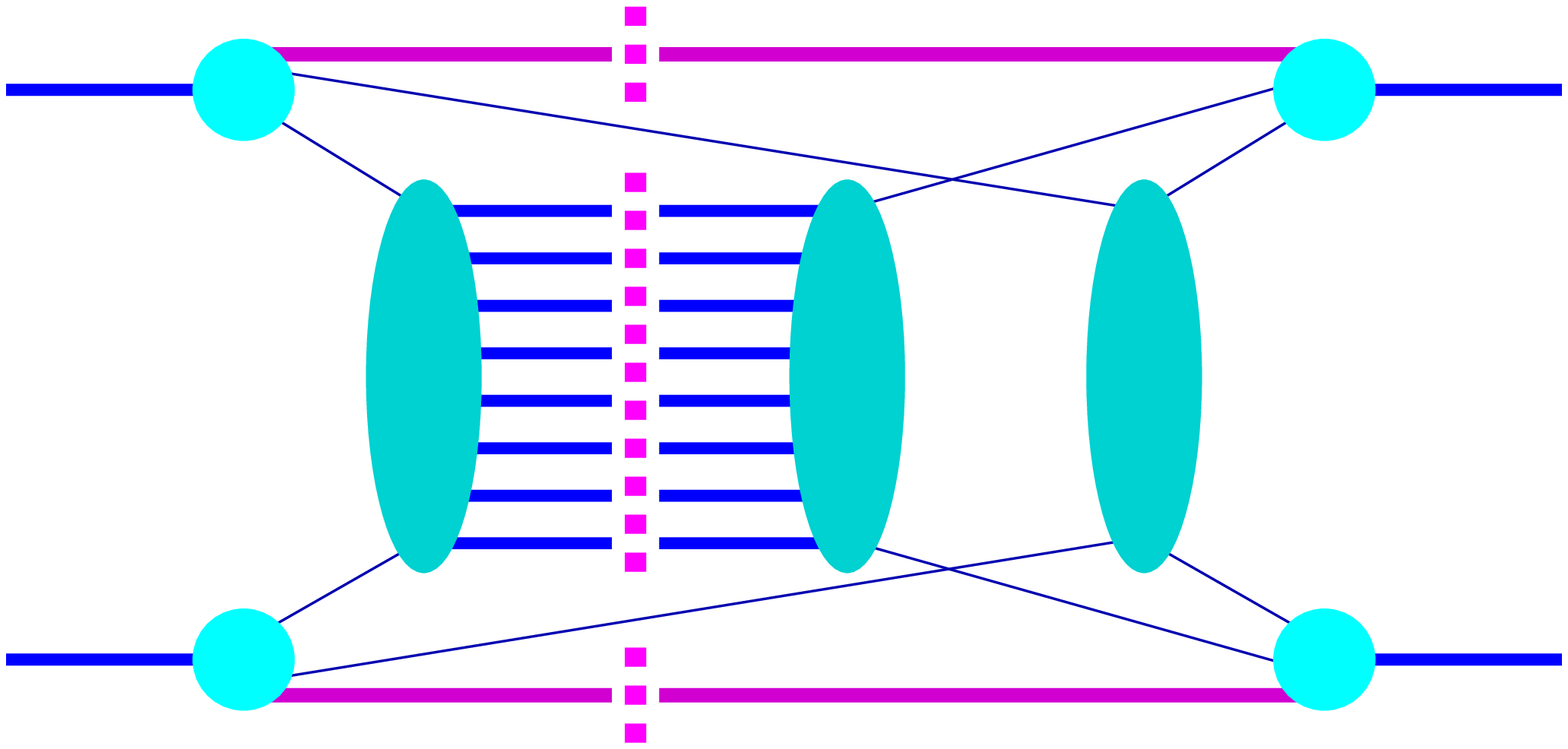}\( \, \,  \)(b)\includegraphics[  scale=0.4]{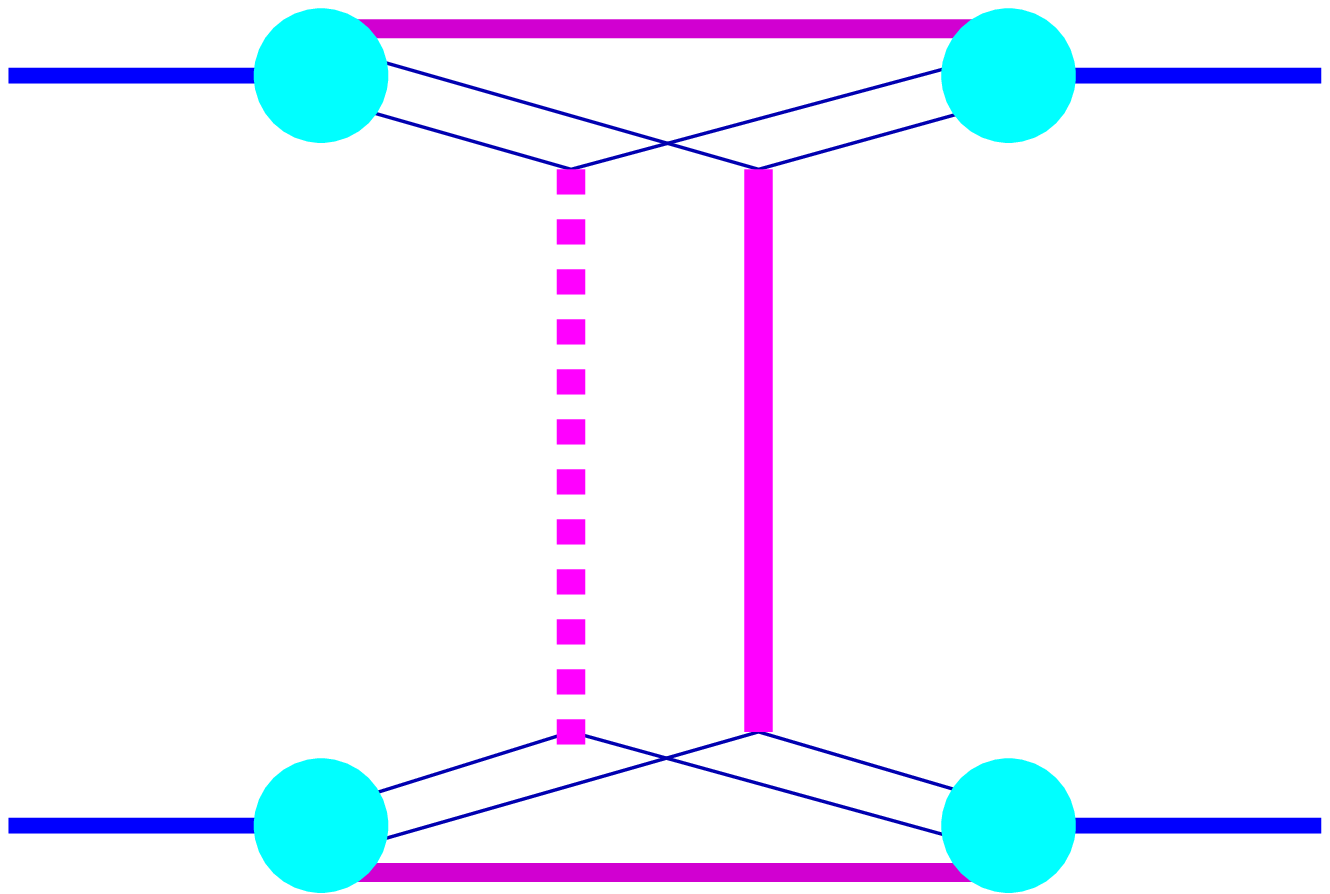}\par}

\caption{Inelastic scattering in pp. a) An interference term of cross section,
b) Represented with our simplified notations.\label{t7b}}
\end{figure}

To calculate cross sections, we need to square the amplitude, which
leads to many interference terms, as the one shown in Fig. \ref{t7b}(a),
which represents interference between the first and the second diagram
of Fig. \ref{t7}. We use the usual convention to plot an amplitude
to the left, and the complex conjugate of an amplitude to the right
of some imaginary {}``cut line'' (dashed vertical line). The left
part of the diagram is a cut elementary diagram, conveniently plotted
as a dashed line, see Fig. \ref{t7b}(b). The amplitude squared is
now the sum over many such terms represented by solid and dashed lines.

\begin{figure}
{\centering \includegraphics[  scale=0.35]{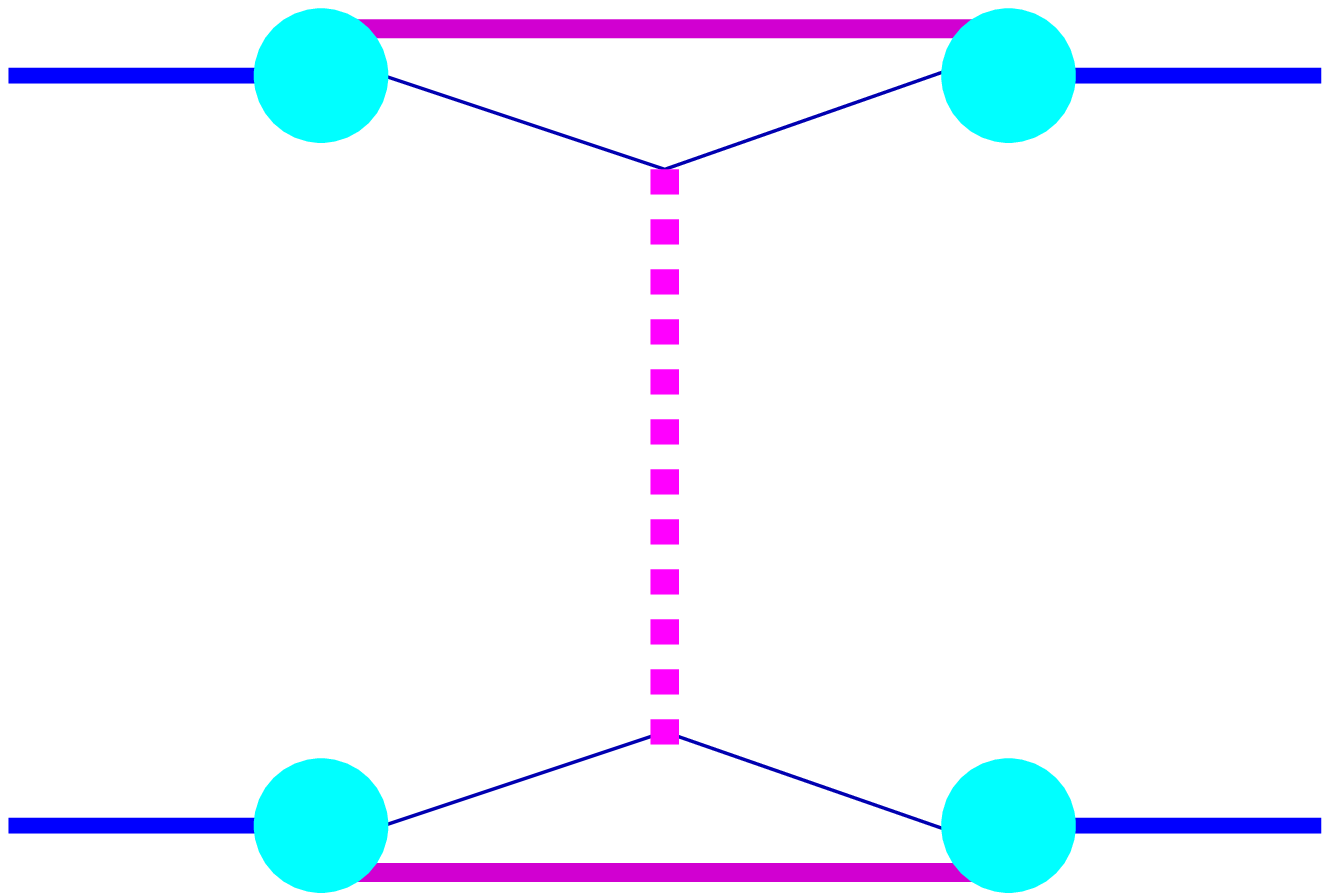}~\( \,  \)\includegraphics[  scale=0.35]{t7ar}\( \,  \)~\includegraphics[  scale=0.35]{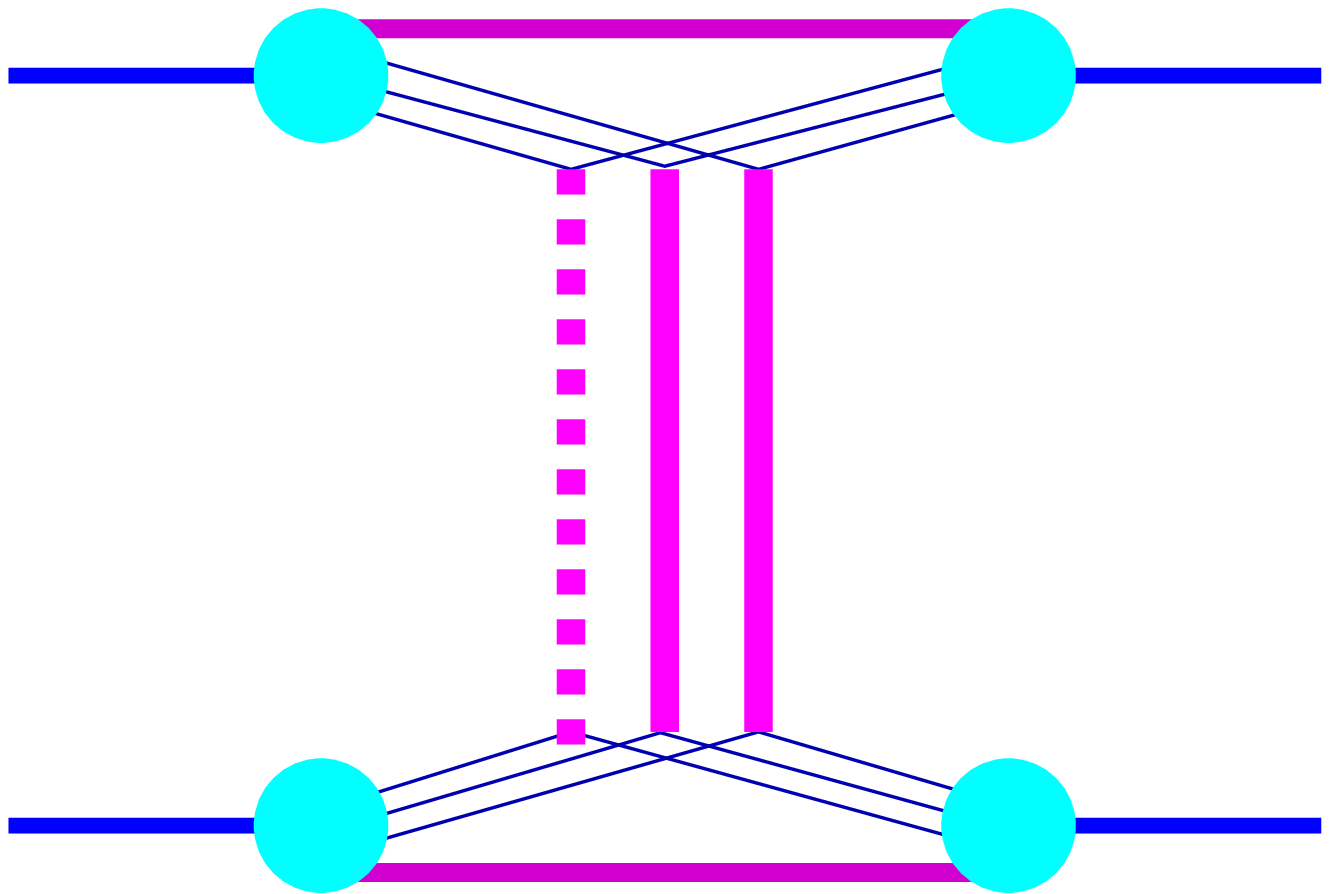}\par}

\caption{Class of terms corresponding to one inelastic interaction.\label{t7c}}
\end{figure}

\begin{figure}
{\centering \includegraphics[  scale=0.35]{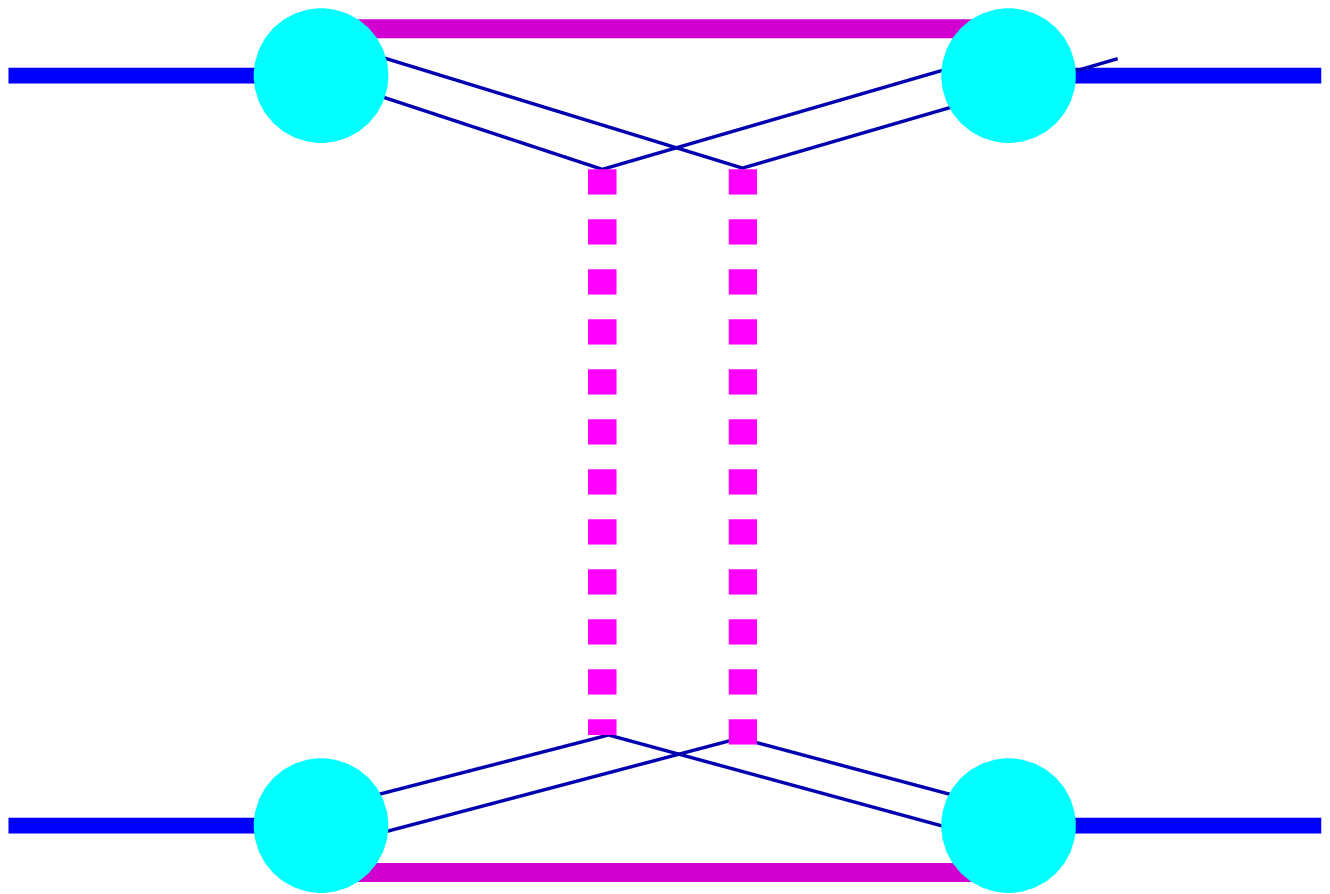}~\( \,  \)\includegraphics[  scale=0.35]{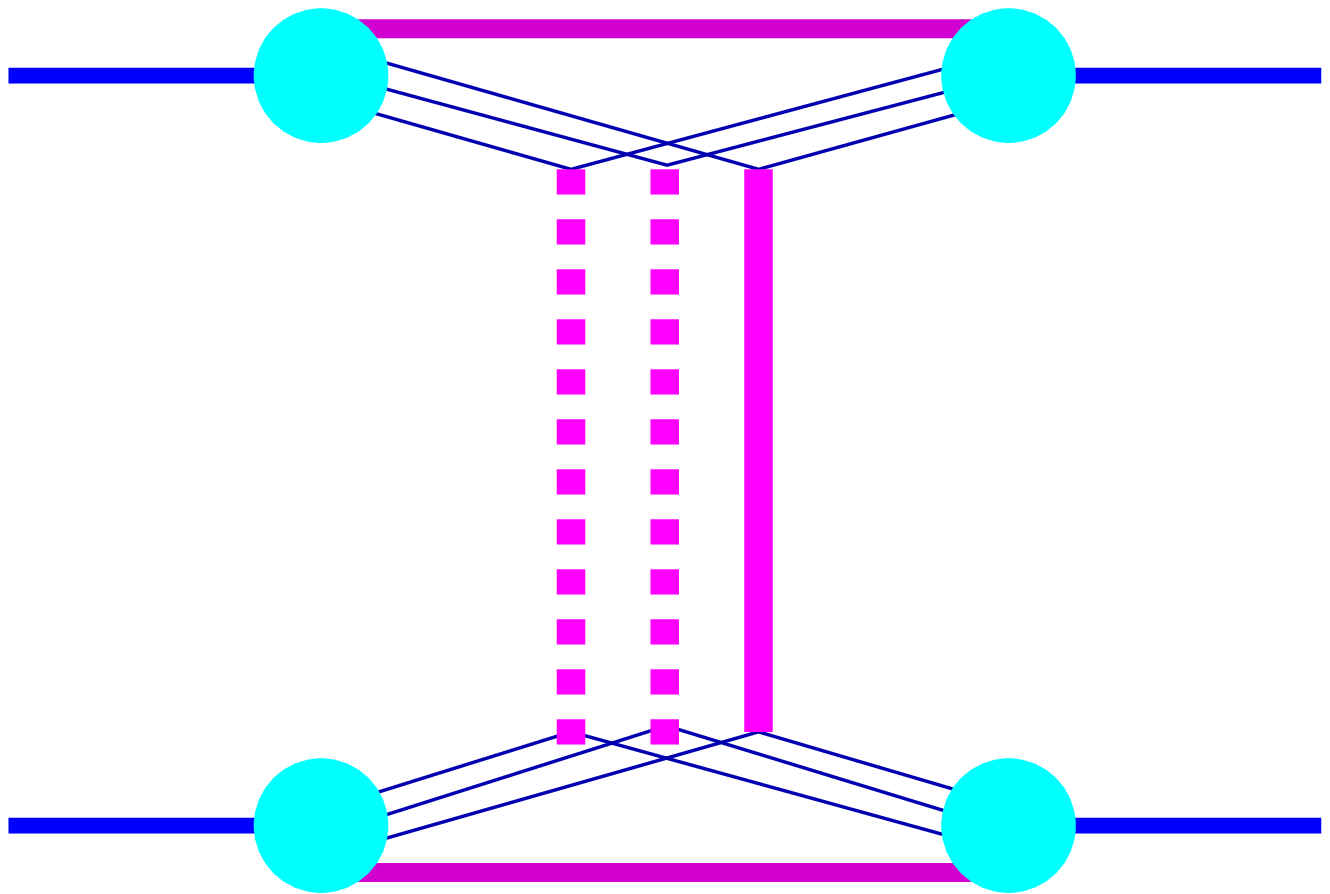}~\( \,  \)\includegraphics[  scale=0.35]{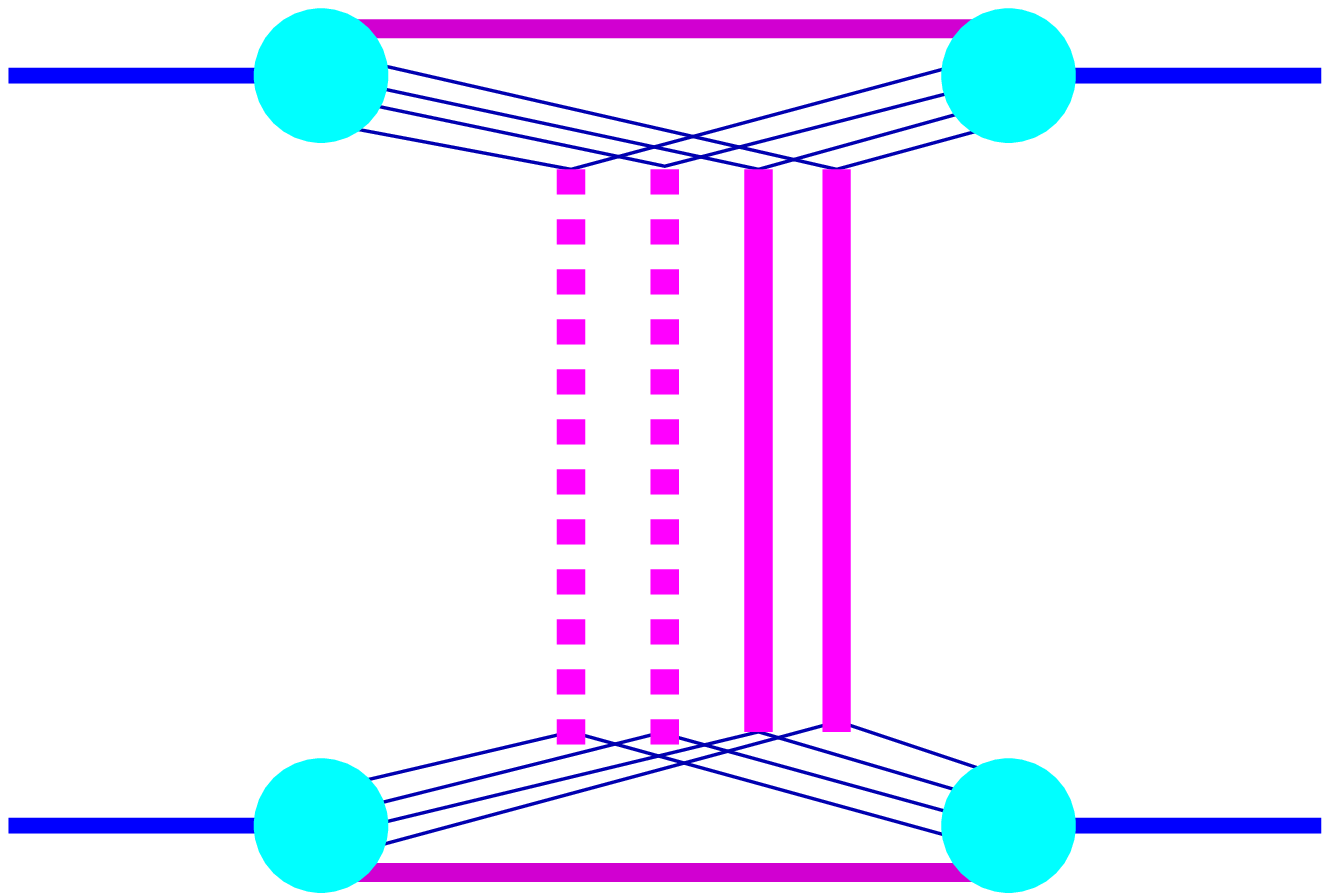}\par}

\caption{Class of terms corresponding to two inelastic interactions.\label{t8c}}
\end{figure}

When squaring the inelastic amplitude, all of the terms which correspond
to the same final state interfere. For example, a single inelastic
interaction does not interfere with a double inelastic interaction,
whereas all the contributions with exactly one inelastic interaction
interfere independent of the number of elastic collisions. So considering
a squared amplitude, one may group terms together representing the
same final state. In our pictorial language, this means that all diagrams
with one dashed line, representing the same final state, may be considered
to form a class, characterized by \( m=1 \) -- one dashed line (
one cut Pomeron) -- and the light cone momenta \( x^{+} \) and \( x^{-} \)
attached to the dashed line (defining energy and momentum of the Pomeron).
In Fig. \ref{t7c}, we show several diagrams belonging to this class,
in Fig. \ref{t8c}, we show the diagrams belonging to the class of
two inelastic interactions, characterized by \( m=2 \) and four light-cone
momenta \( x_{1}^{+} \), \( x_{1}^{-} \), \( x_{2}^{+} \), \( x_{2}^{-} \).

Generalizing these considerations, we may group all contributions
with \( m \) inelastic interactions. The sum of all these terms represents
the probability of having \( m \) inelastic interactions with \( x_{1}^{+} \)..\( x_{2m}^{+} \),
\( x_{1}^{-} \)...\( x_{2m}^{-} \) at a given impact parameter.
Integrating over impact parameter provides the corresponding cross
section. By this we obtain a probability distribution for the number
of elementary interactions (number of Pomerons) and the momenta of
these Pomerons. 
\begin{figure}
{\centering \includegraphics[  scale=0.7]{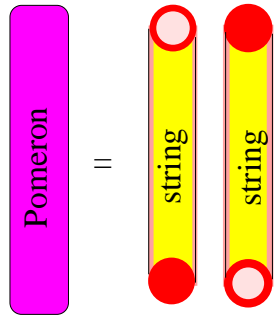}\par}

\caption{Each Pomeron is identified with two strings.\label{pomstri}}
\end{figure}

How to form strings from Pomerons? No matter whether single-Pomeron
or multiple-Pomeron exchange happens in a proton-proton scattering,
all Pomerons are treated identically. Each Pomeron is identified with
two strings, see Fig. \ref{pomstri}.

The string ends are quarks and antiquarks from the sea. This differs
from traditional string models, where all the string ends are valence
quarks. Due to the possibility of having a large number of Pomerons,
this is impossible in our approach. The valence quarks stay in remnants.
Being formed from see quarks, string ends from cut Pomerons have complete
flavour symmetry and produce particles and antiparticles in equal
amounts.

\begin{figure}
{\centering \includegraphics[  scale=0.5]{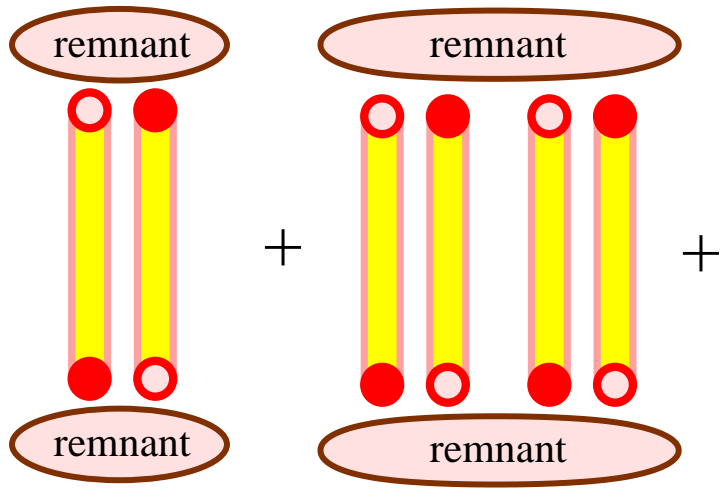}\par}

\caption{Remnants in single (two strings) and double scattering (four strings):
in any case, two remnants contribute.\label{remnstring2}}
\end{figure}

Remnants are new objects, compared to other string models, see Fig.
\ref{remnstring2}.
 The partonic content of a remnant is as follows: three valence quarks
and the corresponding antiparticles of the partons representing the
string ends. The masses of remnants are assumed to be small compared
to the kinetic energies involved and are therefore neglected for the
calculations of multi-Pomeron configurations. To obtained finally
the masses, one parameterize the mass distribution of a remnant as
\( P(m^{2})\propto (m^{2})^{-\alpha } \), \( m^{2}\in (m_{\mathrm{min}}^{2},\, x^{+}s) \),
where \( s \) is the squared energy at center mass system, \( m_{\mathrm{min}} \)
is the minimum mass of hadrons to be made from the remnant's quarks
and antiquarks, and \( x^{+} \) is the light-cone momentum fraction
of the remnant which is determined in the collision configuration.
Through fitting the data at 158 GeV we determine the parameter \( \alpha =1.5 \).
Remnants decay into hadrons according to n-body phase space\cite{droplet}.

\begin{figure}
{\centering \includegraphics[  scale=0.8]{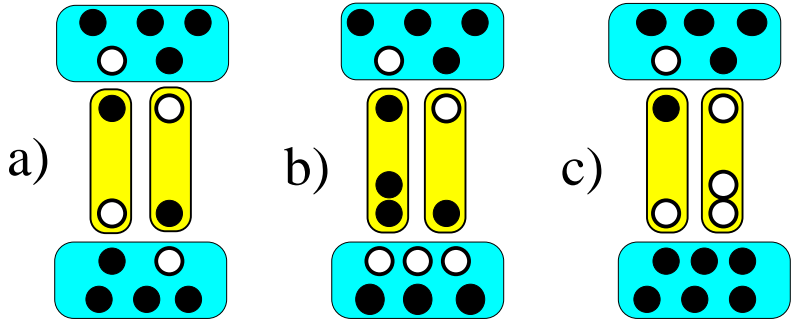}\par}

\caption{\label{nexus2} {\small a) The most simple and frequent collision
configuration has two remnants and only one cut Pomeron represented
by two \protect\( \mathrm{q}-\overline{\mathrm{q}}\protect \) strings.
b) One of the \protect\( \overline{\mathrm{q}}\protect \) string-ends
can be replaced by a \protect\( \mathrm{qq}\protect \) string-end.
c) With the same probability, one of the \protect\( \mathrm{q}\protect \)
string-ends can be replaced by a \protect\( \overline{\mathrm{q}}\overline{\mathrm{q}}\protect \)
string-end. }}
\end{figure}

The leading order and therefore the most simple and most frequent
collision configuration has two remnants and only one cut Pomeron
represented by two \( \mathrm{q}-\overline{\mathrm{q}} \) strings
as in Fig. \ref{nexus2}a. Besides the three valence quarks, each
remnant has additionally a quarks and an antiquark to compensate the
flavour. 

In NE{\large X}US \textbf{3}, this most simple approach is slightly
modified by allowing with a small probability \( P_{qq} \) that an
antiquark \( \bar{\mathrm{q}} \) at one of the legs of the Pomeron
is replaced by a diquark \( \mathrm{qq} \). The corresponding string
ends are then a diquark and a quark. In this way we get quark-diquark
(\( \mathrm{q}-\mathrm{qq} \)) strings from cut Pomerons. The \( \mathrm{qqq} \)
Pomeron end has to be compensated by the three corresponding antiquarks
in the remnant, as in Fig. \ref{nexus2}b. The (\( 3\mathrm{q}3\overline{\mathrm{q}} \))
remnant may decay into three mesons (\( 3\mathrm{M} \)) or a baryon
and an anti-baryon (\( \mathrm{B}+\overline{\mathrm{B}} \)), but
the \( 3\mathrm{M} \) mode is favored by phase space. For symmetry
reasons, the \( \mathrm{q} \) leg of a cut Pomeron is replaced by
an antidiquark \( \overline{\mathrm{q}}\overline{\mathrm{q}} \) with
the same probability \( P_{\mathrm{qq}} \). This yields a \( \overline{\mathrm{q}}-\overline{\mathrm{q}}\overline{\mathrm{q}} \)
string and a (\( 6\mathrm{q} \)) remnant, as shown in Fig. \ref{nexus2}c.
The (\( 6\mathrm{q} \)) remnant decays into two baryons. Since \( \mathrm{q}-\mathrm{qq} \)
strings and \( \overline{\mathrm{q}}-\overline{\mathrm{q}}\overline{\mathrm{q}} \)
strings have the same probability to appear from cut Pomerons, baryons
and antibaryons are produced in the string fragmentation with the
same probability. However, from remnant decay, baryon production is
favored due to the initial valence quarks.

With decreasing energy, the relative importance for the particle production
of the strings decreases as compared to the remnants, because the
energy of the string is lowered as well. If the mass of the string
is lower than the cut off, it will be discarded. However, the fact
that an interaction has taken place is taken into account by the excitation
of the remnant which follows still the above mentioned distribution.

\section{Results}
%After all parameters have been adjusted to the 158 GeV pp reaction
%and the excitation function of the charged particle yield in pp collisions,
%for all identified hadrons at all the other energies and systems,
%the results are predictions. 
To fix the main parameters for the Pomerons and the strings, we use the 
total cross section for pp scattering, the excitation functions for all 
identified charged particle yields from sqrt(s)=10 GeV to 2 TeV and the 
rapidity, pt and multiplicity distributions for at 200 GeV lab and cms 
and 1800 GeV for charged particles. To adjust the parameters of the 
remnant decay and the multistrange baryon production, we mainly use the 
NA49 results at 158 GeV lab. Then, with this set of parameters, 
we compare the simulation results to all pp data at quite wide energy range
and also to np data at beam energy 40 GeV. This comparison work is not trivial, 
firstly because with decreasing beam energy the importance of the remnants
for the particle production increases as compared to the pomerons.
Secondly, pp collisions fix only the sum of the contributions of remnants
and pomerons which contribute to different rapidity regions. The comparison
of the predictions for np collisions with data reveals whether each
individual source of particle production is correctly described. 

\subsection{Hadron Multiplicities}

\subsubsection{Energy Dependence of Average Multiplicities}

\begin{figure}
{\centering \includegraphics[  scale=0.8]{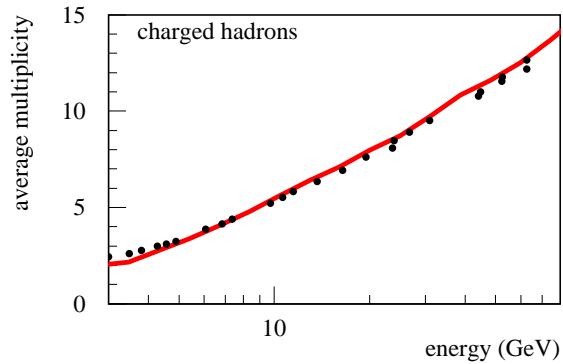}\par}

\caption{\label{multich}Excitation function of charged 
particle \cite{multi}-\cite{multi-ch18} as compared to
the NE{\large X}US results. The $\Lambda$ $\bar \Lambda$ and $K_S$ 
have been identified and their decay products are not included.  }
\end{figure}

We start our investigation with the \( 4\pi  \) multiplicities at different
energies. In fig. \ref{multich} we display the excitation function of 
charged particles \cite{multi}-\cite{multi-ch18} as compared to the NE{\large X}US 3 results.
The $\Lambda$ $\bar \Lambda$ and $K_S$ have been identified
and their decay products are not included here. The agreement between
calculation and experiment shows that this approach is able to describe
excitation functions.

\begin{figure}
{\centering \includegraphics[  scale=0.75]{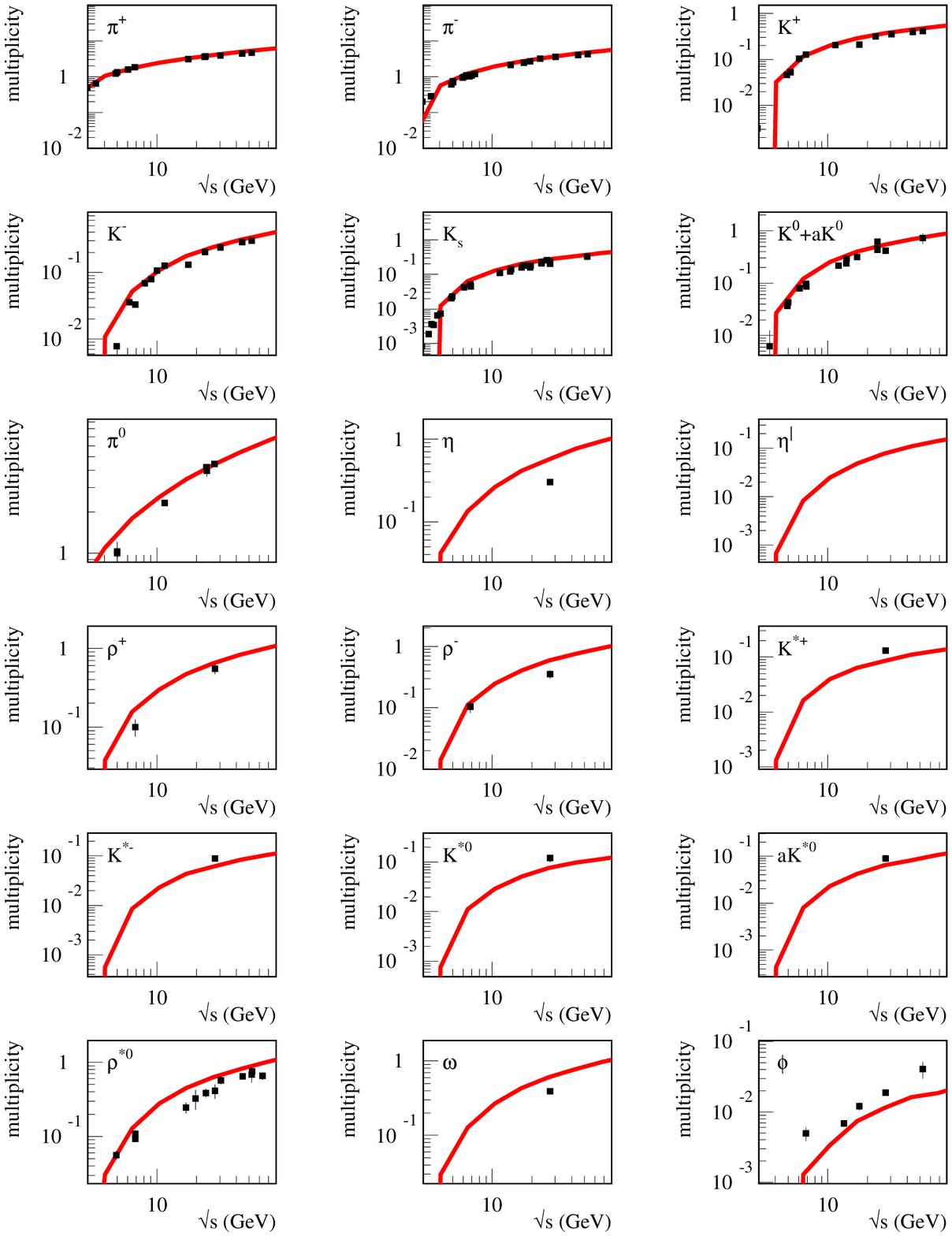}\par}

\caption{\label{multi1}Excitation function of the multiplicity of 
identified hadrons as compared with the available data (squares)
\cite{multi}-\cite{multi4}. 
% Most squared data points are from
%\cite{multi}. Ks, Lambda and antiLambda data are from \cite{multi6},
%data points at \protect\( \sqrt{s}=17.3\protect \)GeV from \cite{multi2},
%at \protect\( \sqrt{s}=27.5\protect \)GeV from \cite{multi4},
% and star data points are from UA5 col. for proton antiproton 
% collsions\cite{multi5}.
 }
\end{figure}

\begin{figure}
{\centering \includegraphics[  scale=0.75]{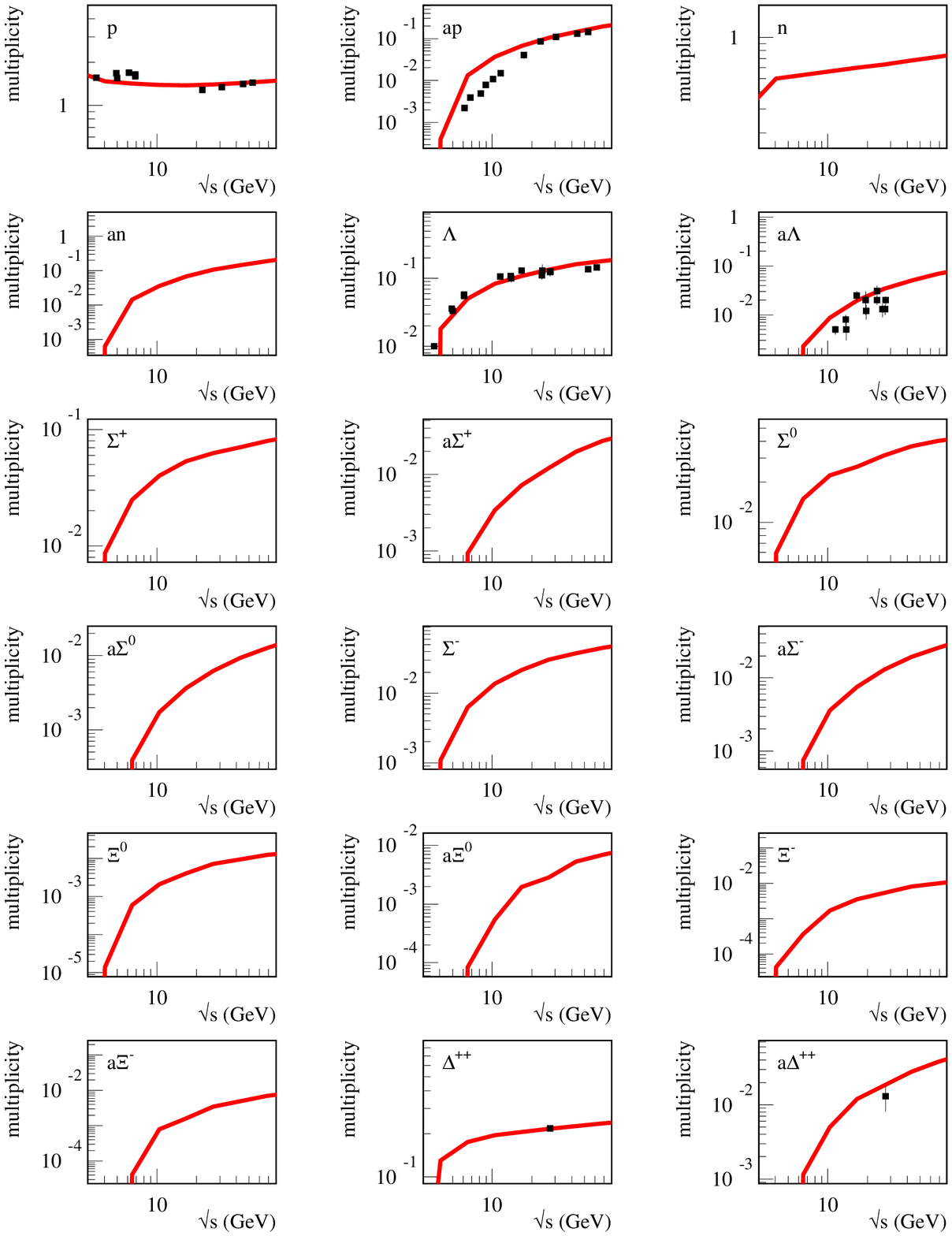}\par}

\caption{\label{multi2} The same as Fig. \ref{multi1}.}
%Triangle points are two times the forward hemisphere yields in NA49 pp 
%experiment \cite{multi7}.  
\end{figure}

\begin{figure}
{\centering \includegraphics[  scale=0.75]{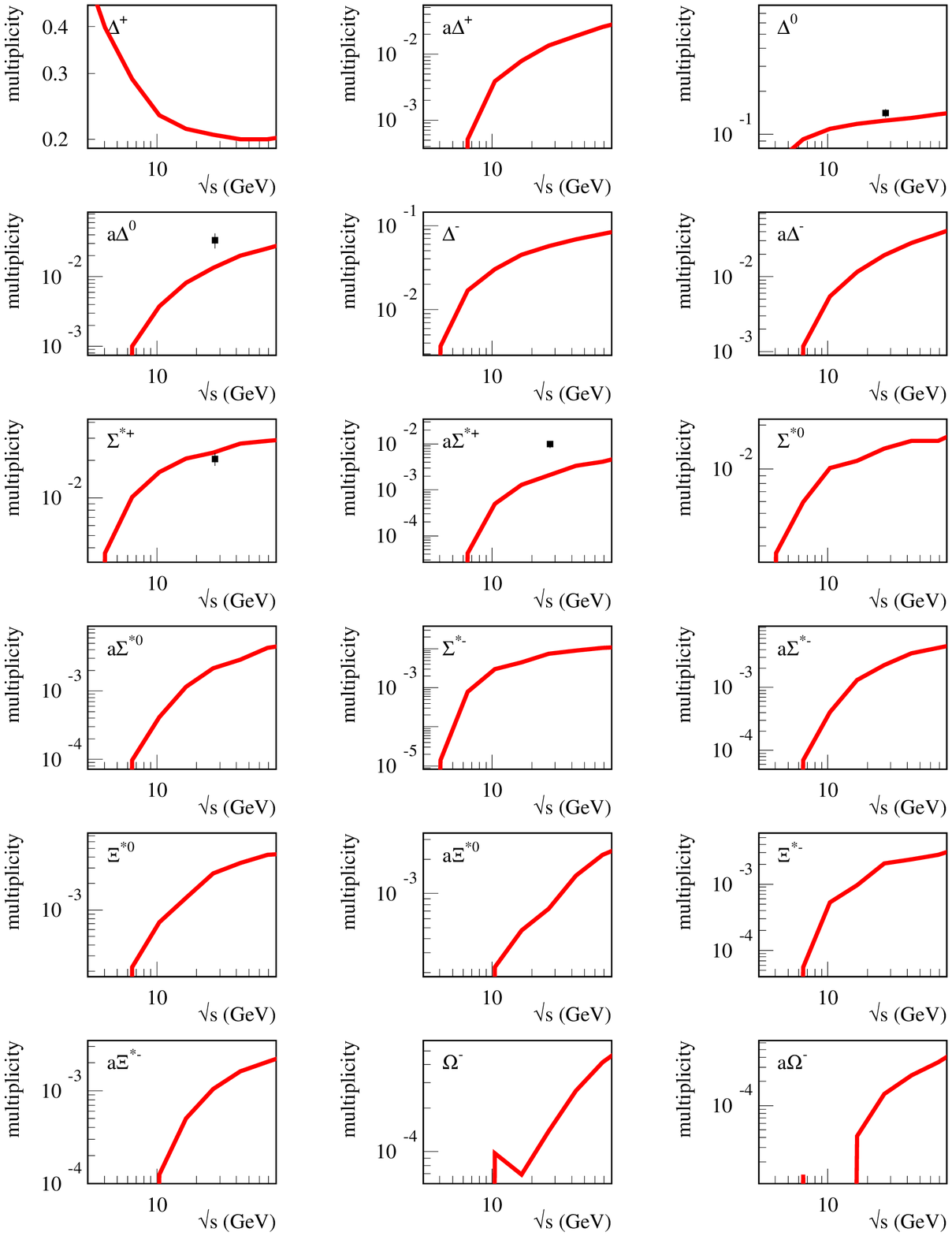}\par}

\caption{\label{multi3} The same as Fig. \ref{multi1}. }
\end{figure}

Now we come to the excitation function of identified hadrons.
The predictions for \( 4\pi  \) multiplicities of identified hadrons 
are shown in figs. \ref{multi1},\ref{multi2},\ref{multi3} and here as well we
find agreement with most of the data over a wide energy range, 
from \( \sqrt{s}=5\, \mathrm{GeV} \) to \( 63\, \mathrm{GeV} \). The largest
discrepancies we observe for the $\phi$ meson and for the $\bar p$ at low
energies. It would be interesting to see in which rapidity region this 
discrepancies appears ( especially because the $\bar \Lambda$ is reasonable well
reproduced but unfortunately such data are not available.

\subsubsection{$K/\pi$ ratios}
\begin{figure}
{\centering \includegraphics[  scale=0.75]{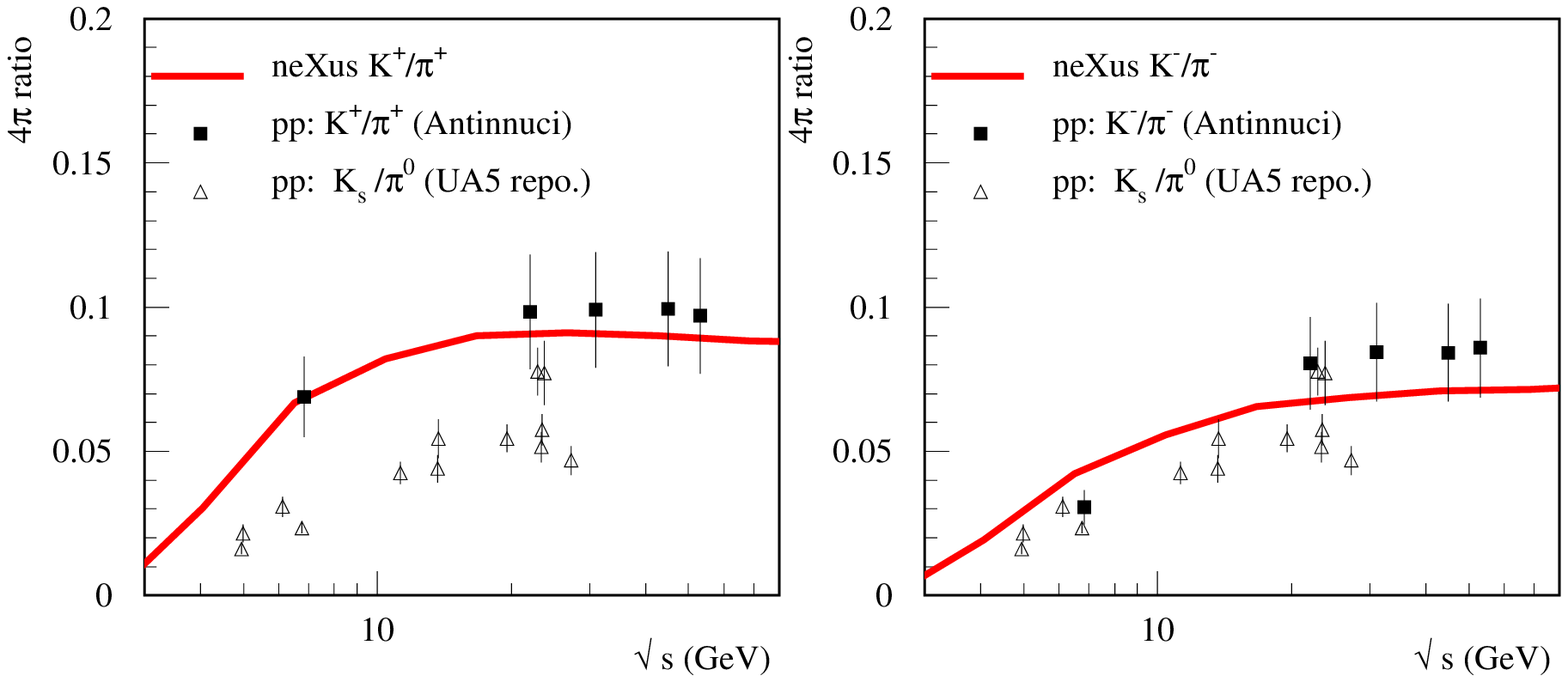}\par}

\caption{\label{kpi} Theoretical and experimental $K/\pi$ ratio. 
Squared points are from \cite{Antinucci}, 
triangle points represent the $K_s/\pi_0$ ratio and are from 
\cite{UA5repo}-\cite{Drijard:1981wg}, compiled in \cite{multi}. }
\end{figure}

Special interest has recently gained the excitation function of the $K/\pi$
ratio because it has been claimed that the appearance of a maximum in the
$K^+/\pi^+$  excitation function, observed in heavy ion reactions, 
may be a signature that the plasma of quarks and gluons is formed. 
Although we are not concerned with heavy ion 
reactions here we display these ratios for later reference in fig. \ref{kpi}.
Neither data nor calculation show a maximum of the $K^+/\pi^+$ ratio in this
elementary collision.

\subsubsection{Rapidity Spectra}

\begin{figure}
{\raggedright \includegraphics[  scale=0.85]{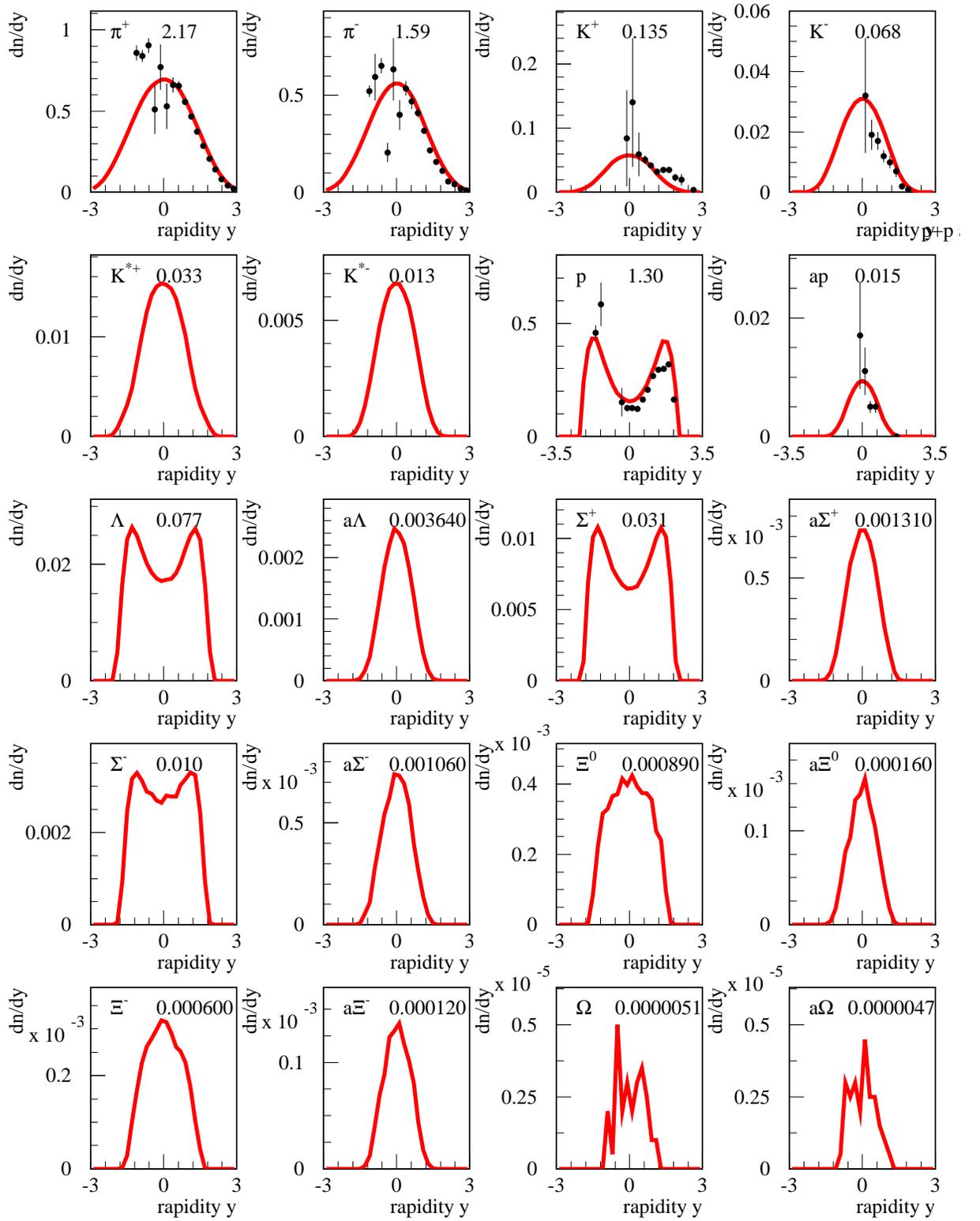}\par}

\caption{\label{pp40}Rapidity spectra (lines) and \protect\( 4\pi \protect \)
multiplicities (numbers) of identified hadrons from proton-proton
at beam energy 40 GeV. Data are from \cite{na49}. }
\end{figure}

\begin{figure}
{\raggedright \includegraphics[  scale=0.85]{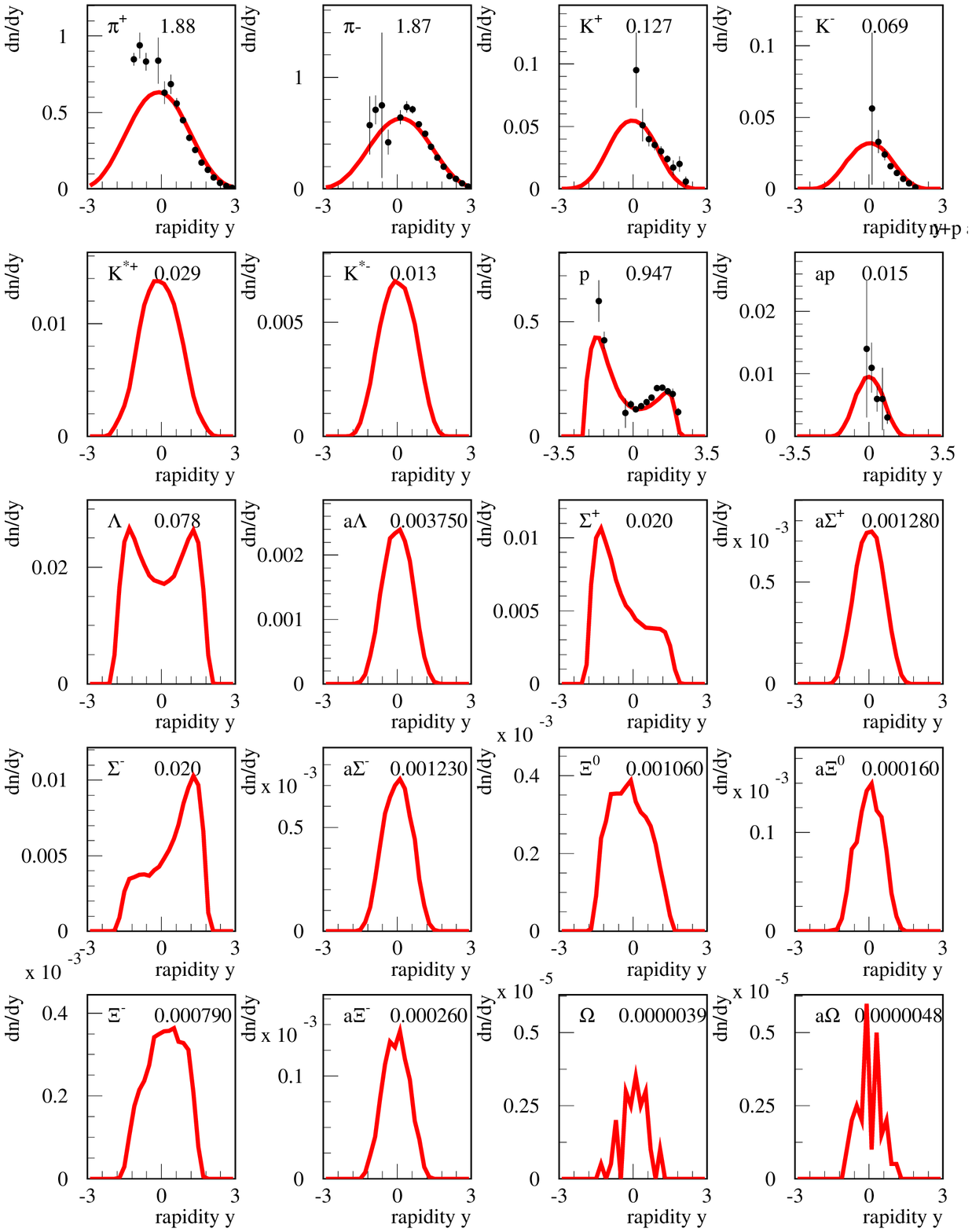}\par}

\caption{\label{np40} Rapidity distribution (lines) and \protect\( 4\pi \protect \)
multiplicities (numbers) of identified hadrons for neutron-proton collisions
at a beam energy of 40 GeV. Data are from \cite{na49}. }
\end{figure}

We come now to more detailed information by studying
the rapidity distribution. Fig. \ref{pp40} shows the rapidity
distribution of a multitude of hadrons for the reaction 40 GeV pp.
Where data from the NA49 collaboration are available we have included
them in the plot. The numbers give the average multiplicity of the
hadrons in 4\( \pi  \). We see that the experimental data
are reasonable well described. The non-strange baryons as well as
those which contain one strange quark show a double hump structure,
the others are peaked at midrapidity. This is a consequence of the
three source structure (two remnants and Pomerons) in our approach.
The leading baryon has still the quantum number of the incoming baryon
but is moderately excited. Therefore it may disintegrate into baryons
whose quantum numbers differ not too much. Also the calculation of
the np reaction at 40 GeV, displayed in Fig. \ref{np40}, reproduces
the data quite nicely. We see of course a much lower number of protons
in the fragmentation region. This difference to the pp collision is
well reproduced. This agreement validates the correct description
of the basic mechanism for particle production in the fragmentation
region. A comparison between the spectra shows that the difference
between pp and np in the pion spectra extends to negative rapidities
(as seen as well in the data). For \( \Lambda  \) and \( \bar{\Lambda } \)
as well as for the other charge zero particles, the difference between
pp and np is negligible. Naturally \( \Sigma ^{+} \)(\( \Sigma ^{-} \))
are more copiously produced if the projectile is a p(n). 

\begin{figure}
{\raggedright \includegraphics[  scale=0.85]{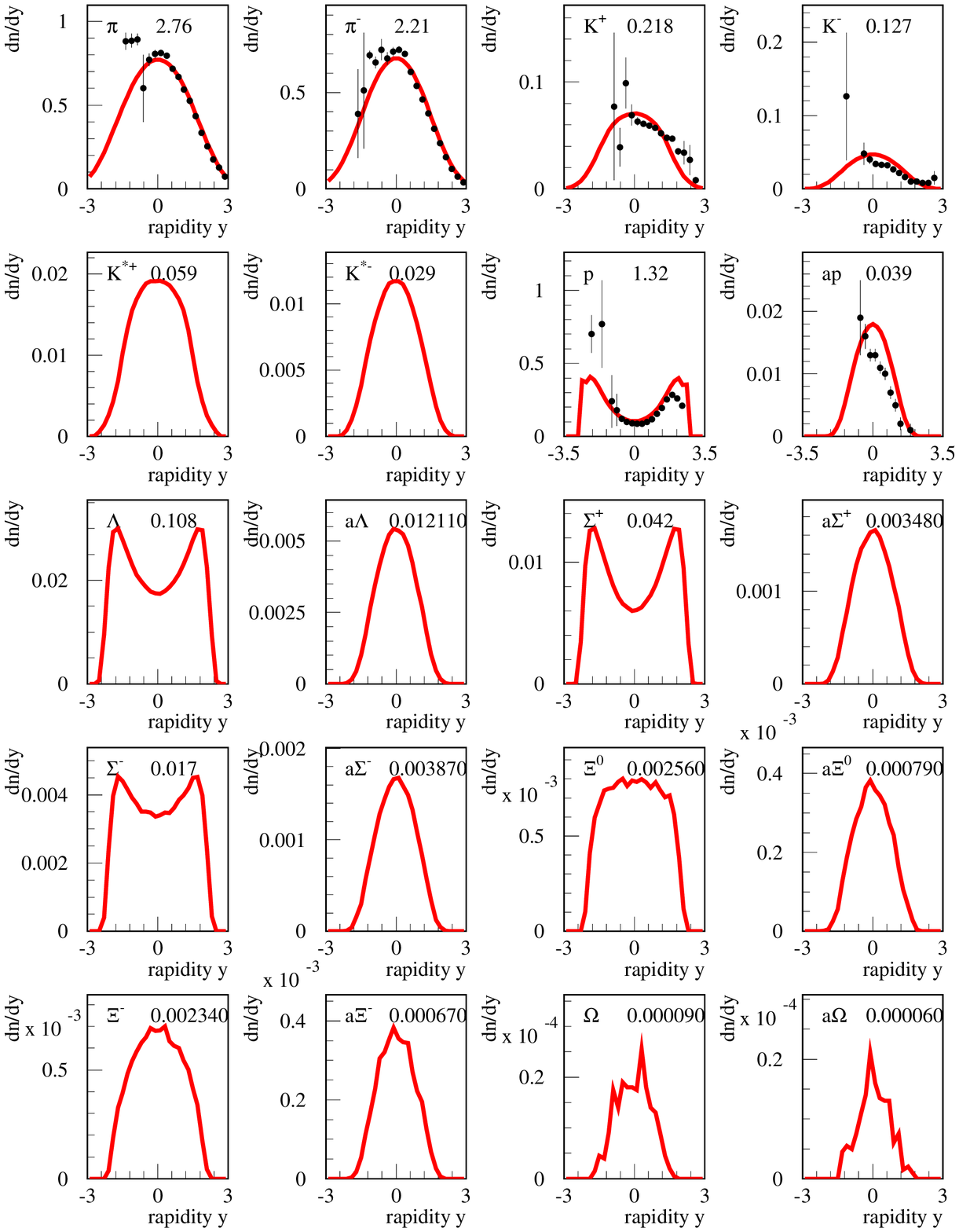}\par}

\caption{\label{pp100} Rapidity distribution (lines) and \protect\( 4\pi \protect \)
multiplicities (numbers) of identified hadrons for proton-proton collisions
at a beam energy of 100 GeV. Data are from \cite{na49}. }
\end{figure}

\begin{figure}
{\raggedright \includegraphics[  scale=0.85]{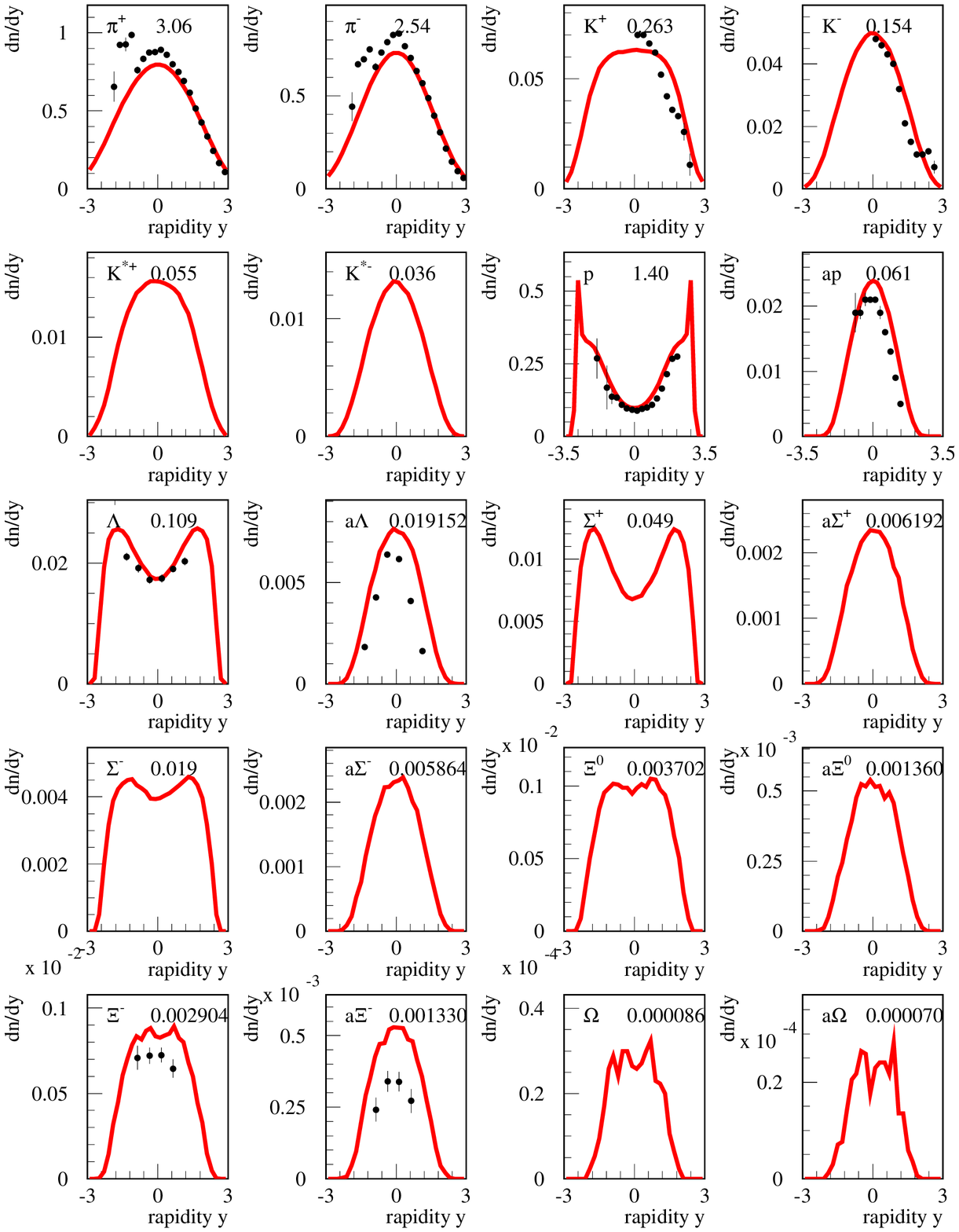}\par}

\caption{\label{pp158}Rapidity distribution (lines) and \protect\( 4\pi \protect \)
multiplicities (numbers) of identified hadrons for proton-proton collisions
at a beam energy of 158 GeV. Data are from \cite{na49}. }
\end{figure}

\begin{figure}
{\raggedright \includegraphics[  scale=0.85]{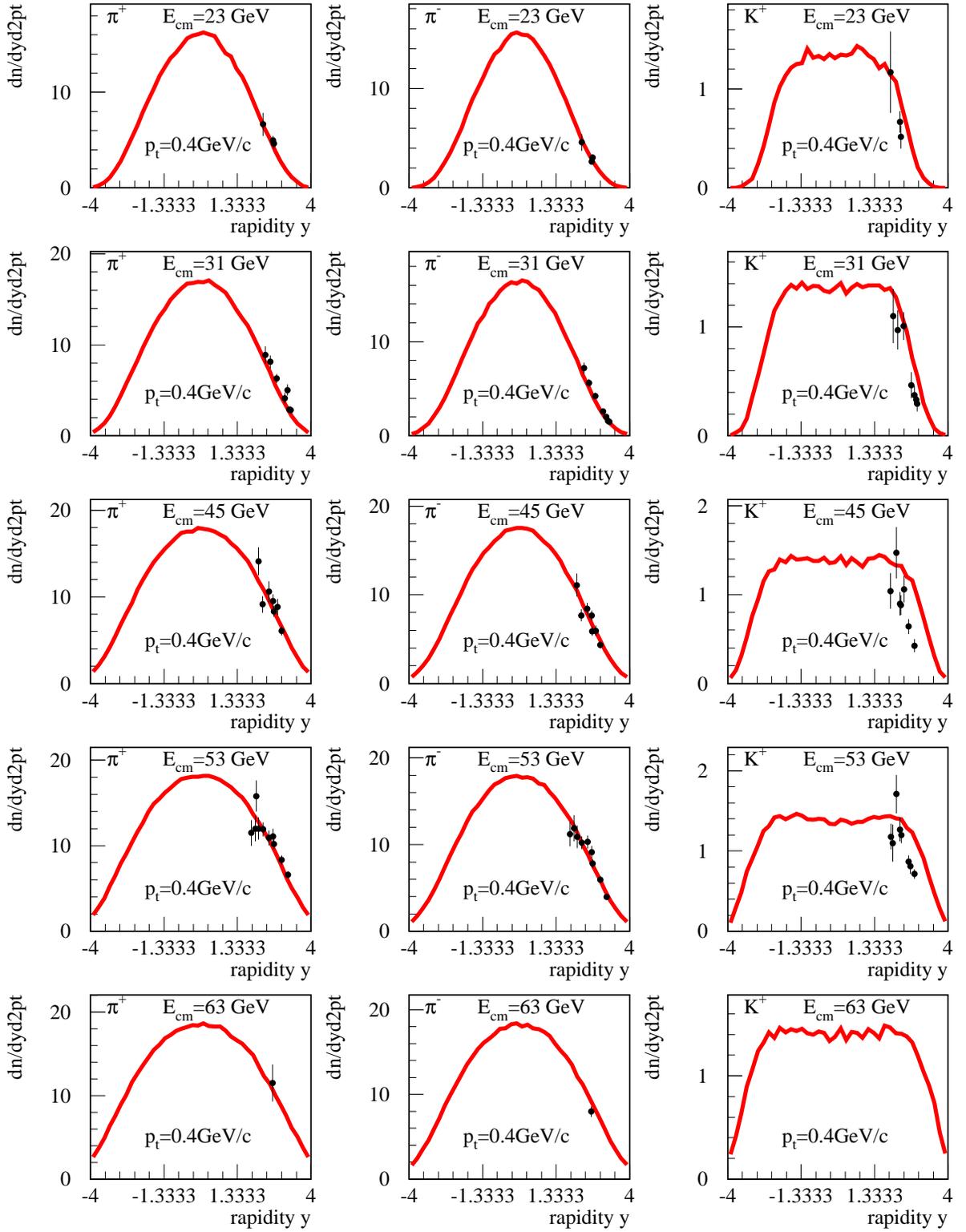}\par}

\caption{\label{isrp1} \protect\( \pi ^{+}\protect \), 
\protect\( \pi ^{-}\protect \)
and \protect\( K^{+}\protect \) rapidity distributions 
at \protect\( p_{t}\protect \)=0.4GeV/c
for proton-proton collisions at ISR energies. 
Data are from \cite{yptspectra}.}
\end{figure}

\begin{figure}
{\raggedright \includegraphics[  scale=0.85]{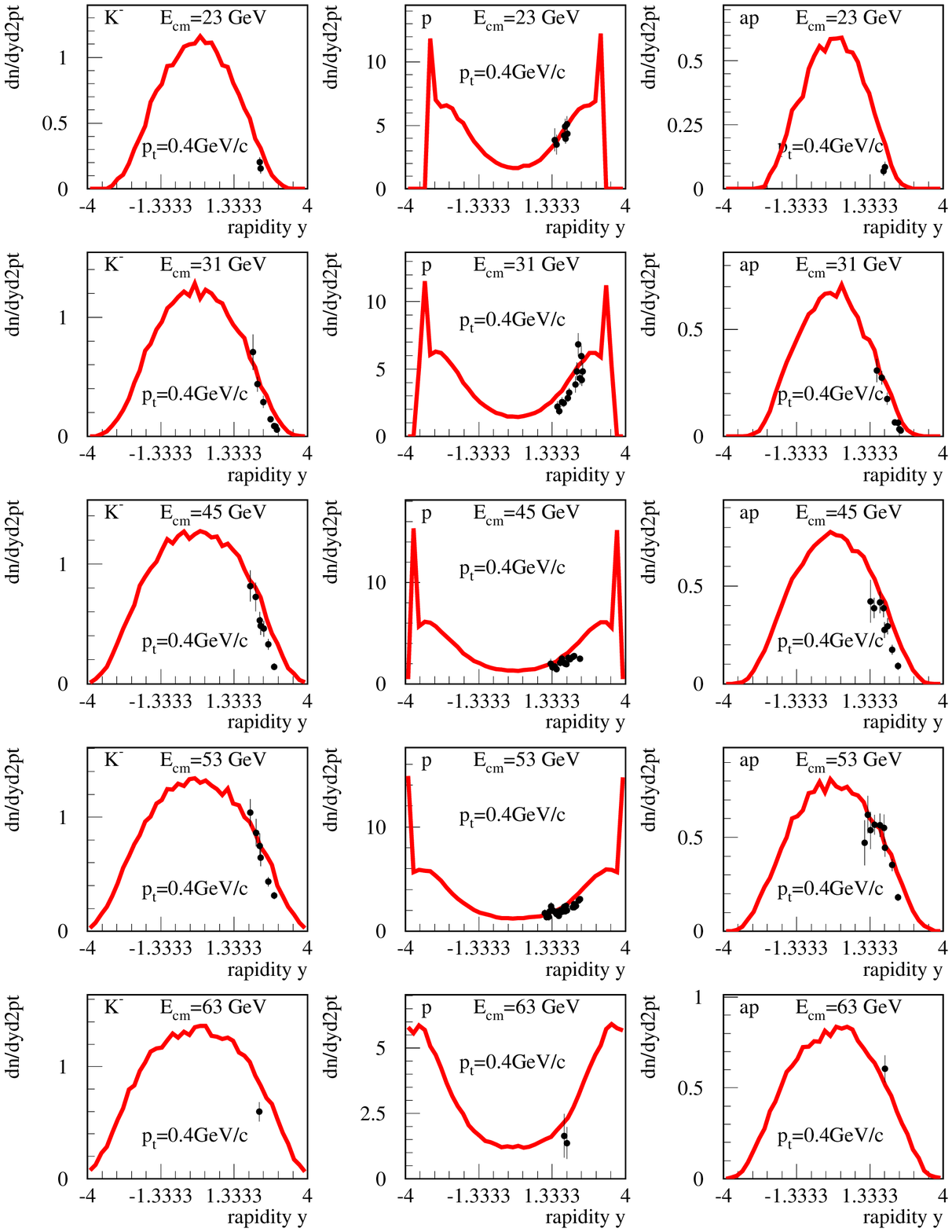}\par}

\caption{\label{isrp2} \protect\( K^{-}\protect \), proton and antiproton
rapidity distributions at \protect\( p_{t}\protect \)=0.4GeV/c for 
proton-proton collisions at ISR energies. Data are from \cite{yptspectra}. }
\end{figure}

Fig. \ref{pp100} shows the rapidity spectra for pp at 100 GeV. Again
we see a quite reasonable agreement between data and calculation. 
At 158 GeV, Fig. \ref{pp158}, we included the experimental \( \Lambda  \),
\( \bar{\Lambda } \), \( \Xi  \), \( \bar{\Xi } \) spectra which
have recently been published \cite{nexn}. We observe also more \( \Omega  \)
than \( \bar{\Omega } \) as seen in experiment \cite{na49}. This
is a consequence of the modification of NE{\large X}US 3 explained
in \cite{nexn} as compared to the original NE{\large X}US 2 version
\cite{nexo} which yields more \( \bar{\Omega } \) than \( \Omega  \)
due to the string topology .

For ISR energies there are only rapidity distributions for a given transverse
momentum. In figs. \ref{isrp1} and \ref{isrp2} we display for \( p_{T} \)
= 0.4 GeV/c the rapidity distributions for \( \pi ^{+},\, \pi ^{-},
\,K^{+},\,K^{-},\,p,\,\bar{p} \)
for energies in between \( E_{cm} \) = 23 GeV and 63 GeV. We see
that also here the spectra agree well with the data where data are
available.

\begin{figure}
{\raggedright \includegraphics[  scale=0.85]{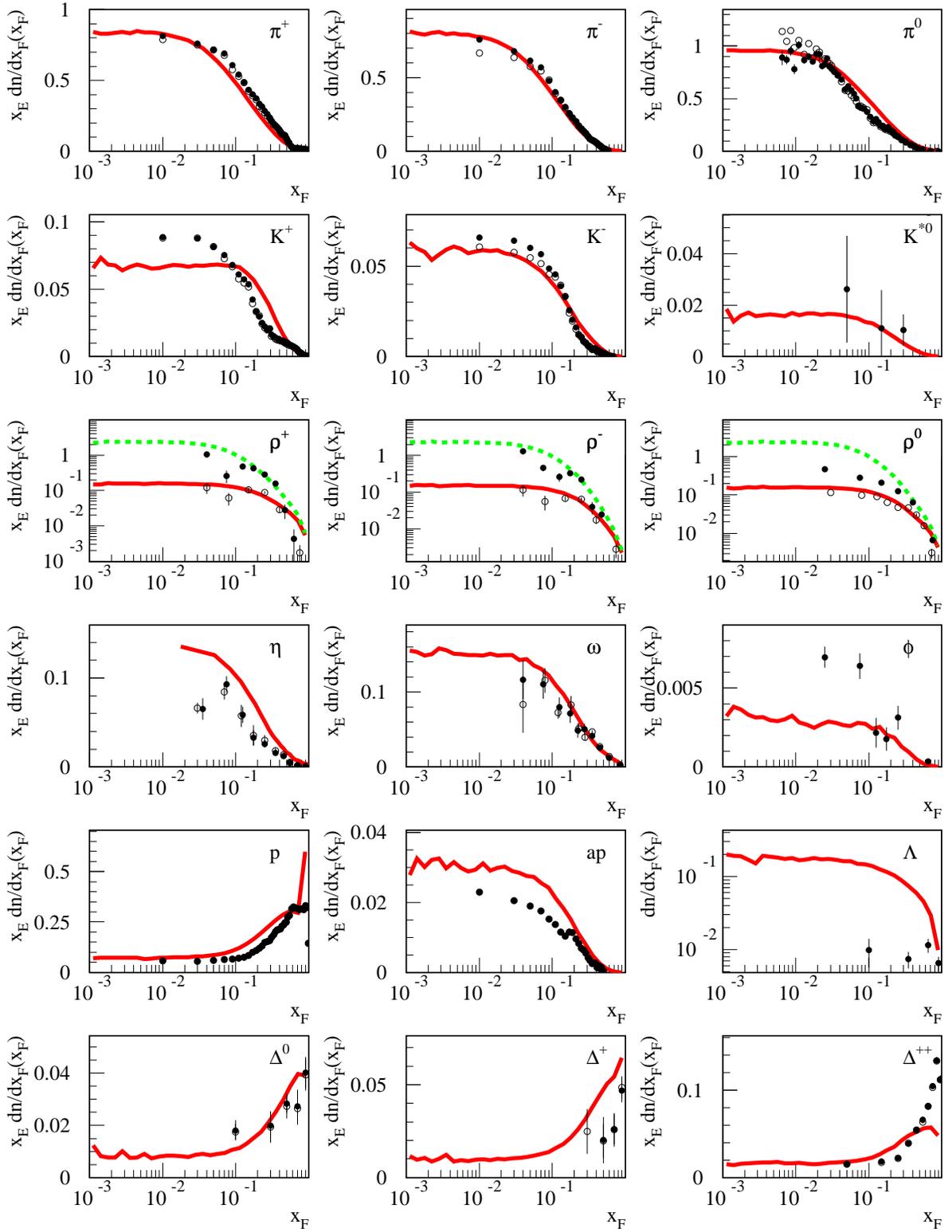}\par}

\caption{\label{pp400} \protect\( x_{F}\protect \) distributions of identified
hadrons from pp collisions at beam energy 400 GeV. Data are from \cite{multi4}.
The vertical axis of dashed lines and the points in rho meson plots is 
\protect\( dn/dx_{F}\protect \). }
\end{figure}

Fig. \ref{pp400} shows the longitudinal \protect\( x_{F}\protect \) distributions of identified
hadrons from pp collisions at a beam energy 400 GeV.  In order
to be comparable with the before-mentioned rapidity distributions, we use a
logarithmic representation of the x-axis. We see
that also here the spectra agree reasonably well with the  LEBC-EHS Col. 
data\cite{multi4}.
Please note that there are two curves for the $\rho$'s. This is due to the
inconsistency of the two available data sets: 
$x_{E} \ dn/dx_{F}$ and $dn/dx_{F}$ spectra. So $dn/dx_{F}$ spectra from 
our simulations are additionally plotted as dashed lines there.

\subsection{Transverse Momentum}

\subsubsection{Average Transverse Momenta }

\begin{figure}
{\centering \resizebox*{0.95\textwidth}{!}{\includegraphics{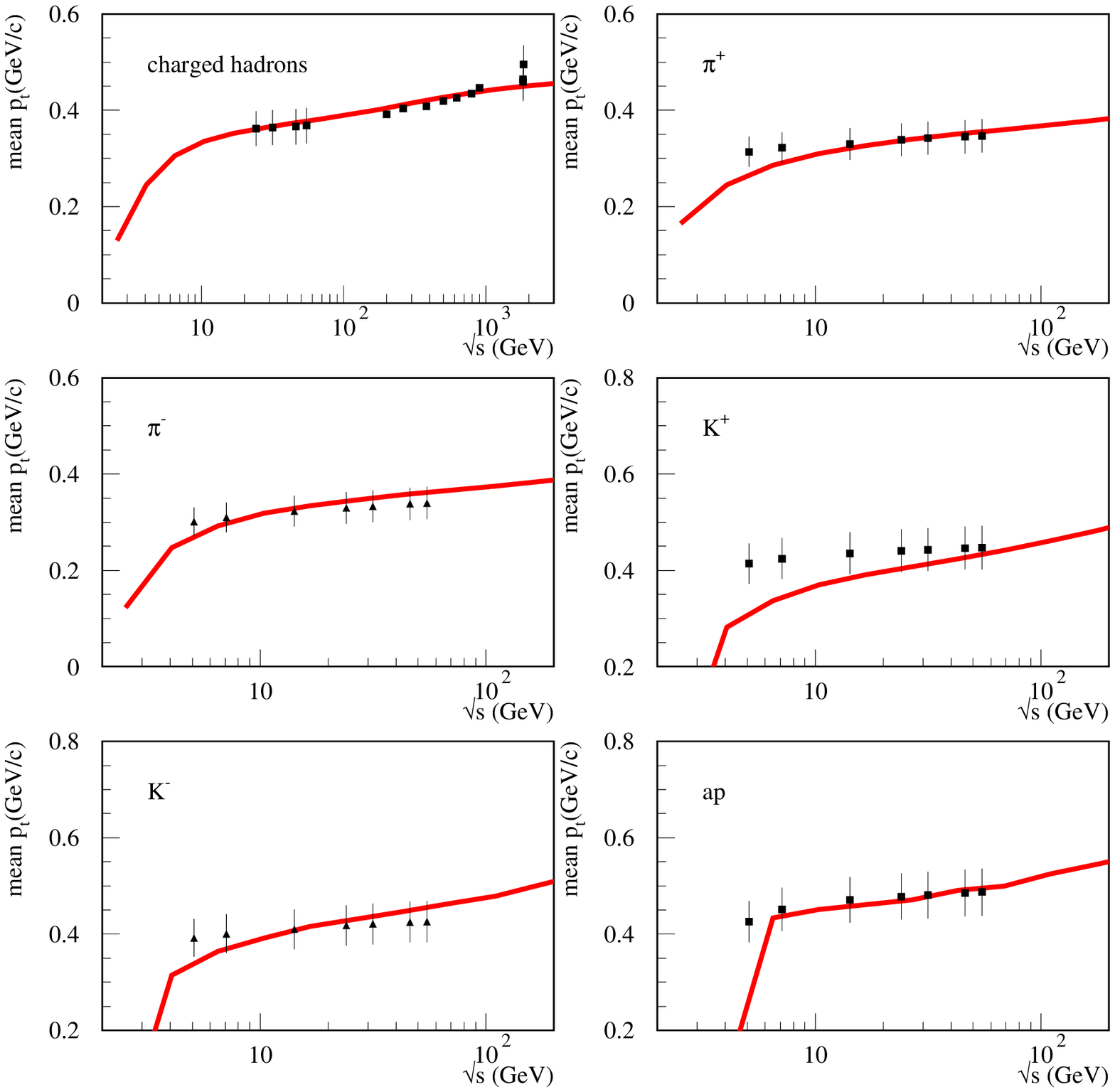}} \par}

\caption{\label{meanpt}Average transverse momentum of charged hadrons, pions,
kaons and antiproton from 4\protect\( \pi \protect \) phase space
at different energies. Data are from \cite{meanpt}.}
\end{figure}

Fig. \ref{meanpt} displays the excitation function of the average
transverse momentum for all charged particles and several particle
species in comparison with the experimental data\cite{meanpt}. 
We see that with the exception of K\( ^{+} \), the average transverse 
momentum is reproduced over the whole kinematical range. For the \( K^{+} \)
we underpredict the average transverse momentum at smaller beam energies.

\subsubsection{Transverse Momentum Distribution}

\begin{figure}
\includegraphics[  scale=0.75]{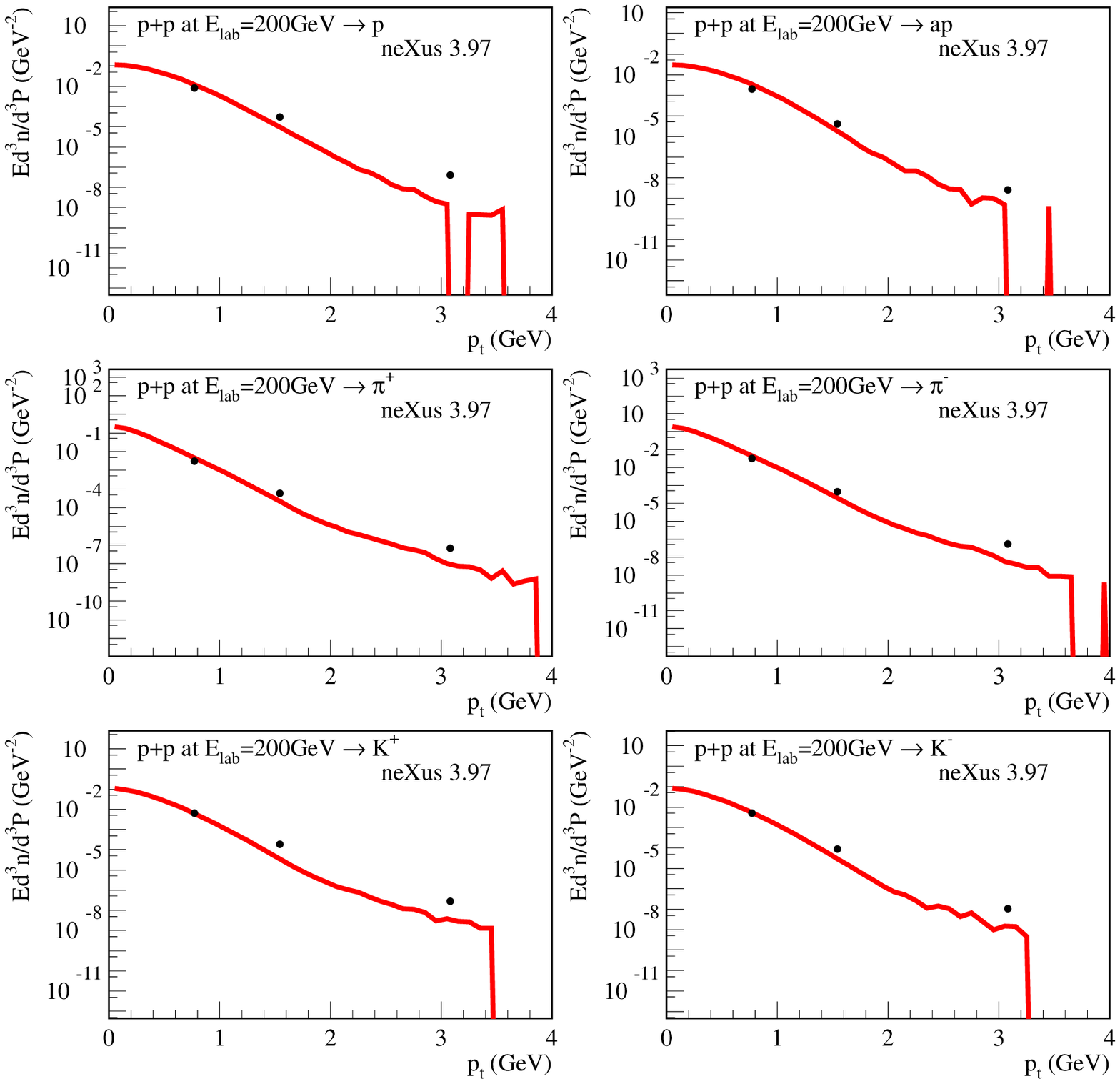}

\caption{Transverse momentum spectra of proton, antiproton, charged pions and
Kaons from pp collisions at \protect\( E_{\mathrm{lab}}=200\mathrm{GeV}\protect \).
Data are from \cite{ptspectra}. \label{pt1}}
\end{figure}

\begin{figure}
\includegraphics[  scale=0.75]{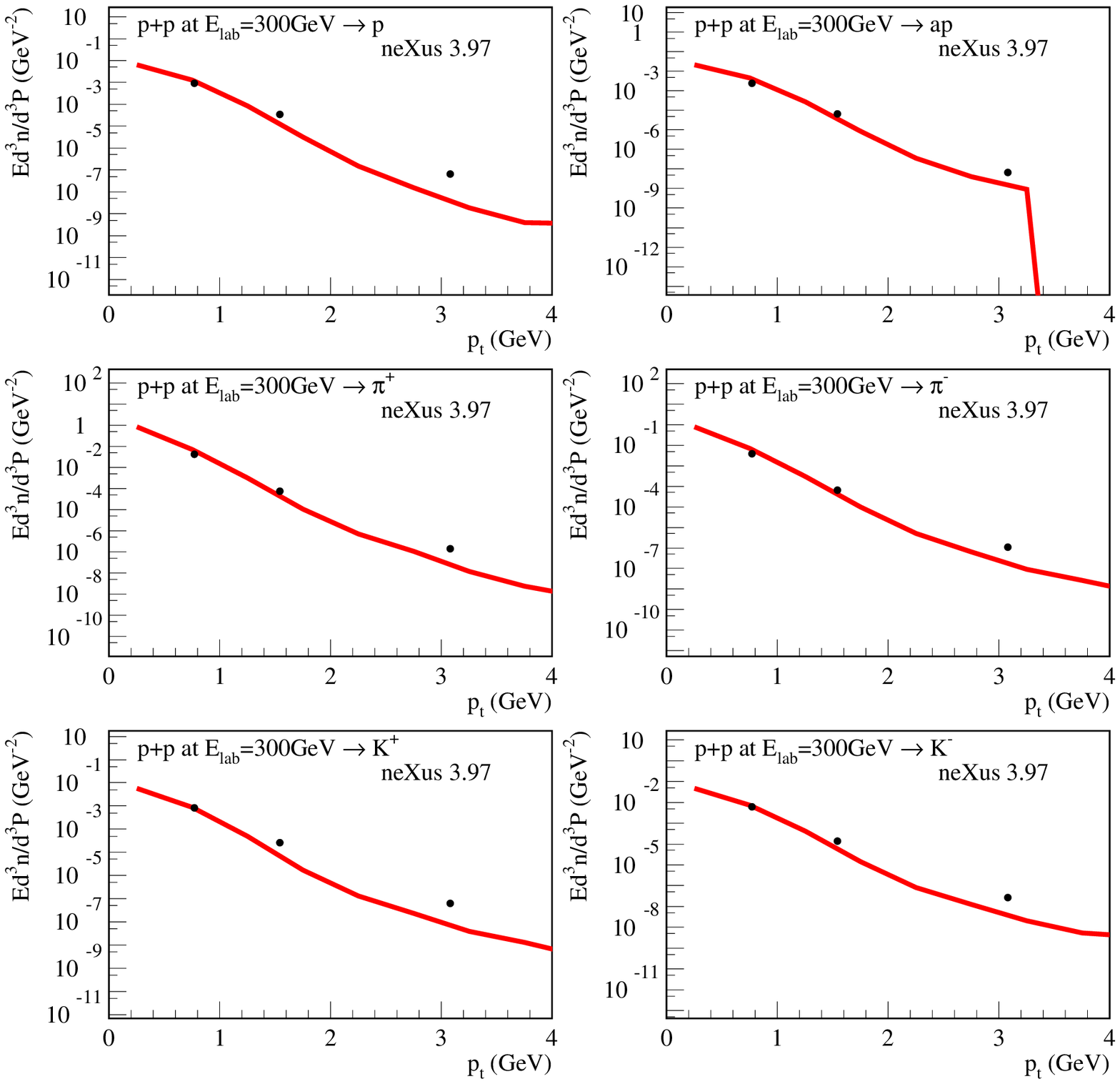}

\caption{Transverse momentum spectra of proton, antiproton, charged pions and
Kaons from pp collisions at \protect\( E_{\mathrm{lab}}=300\mathrm{GeV}\protect \).
Data are from \cite{ptspectra}.\label{pt2}}
\end{figure}

\begin{figure}
\includegraphics[  scale=0.75]{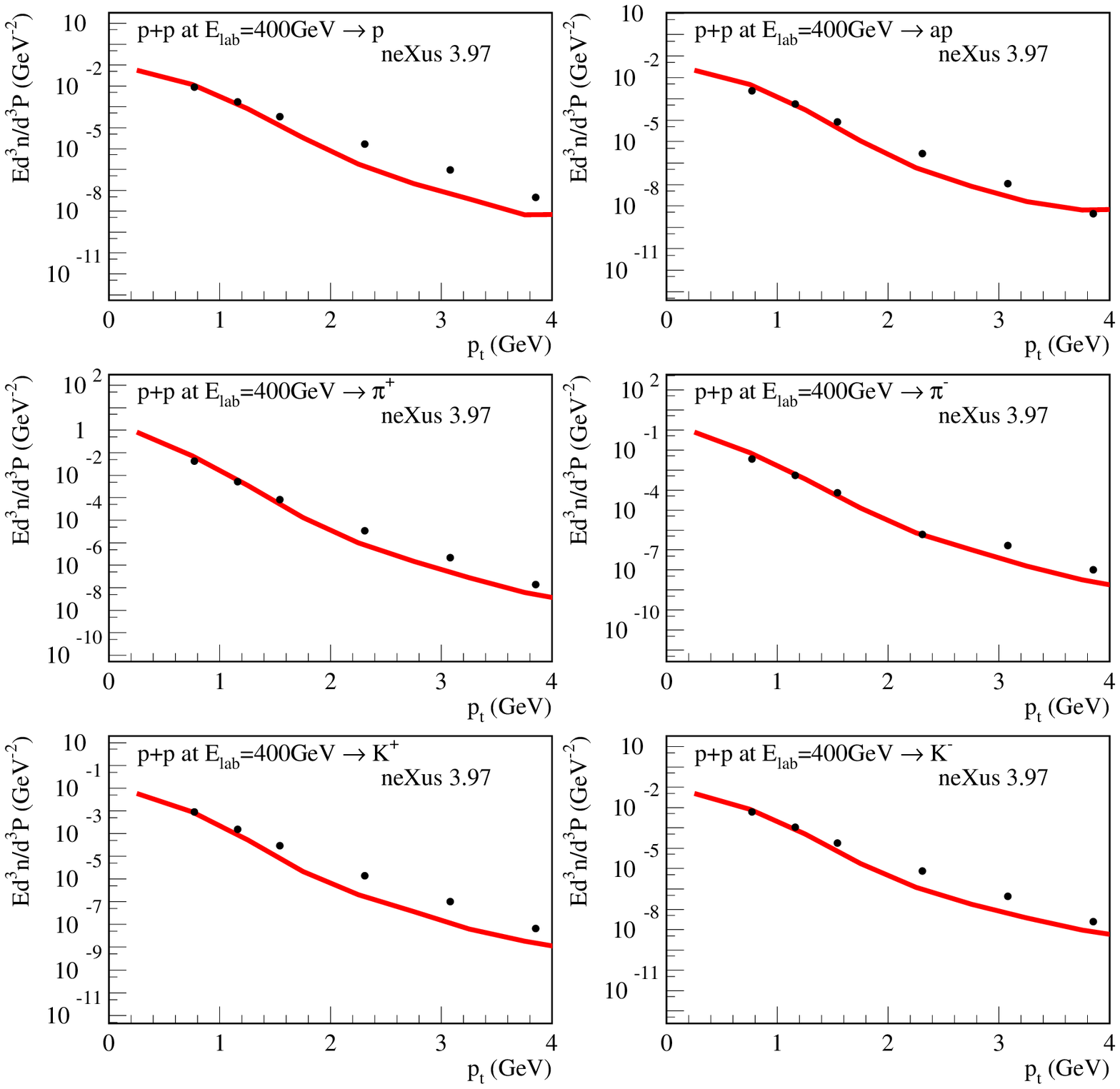}

\caption{Transverse momentum spectra of proton, antiproton, charged pions and
Kaons from pp collisions at \protect\( E_{\mathrm{lab}}=400\mathrm{GeV}\protect \).
Data are from \cite{ptspectra}.\label{pt3}}
\end{figure}

At SPS energies not only the mean transverse momenta but also the
whole transverse momentum spectra is available and we compare these
data with the NE{\large X}US 3 predictions in figs. \ref{pt1},\ref{pt2},\ref{pt3}.
Again, the agreement is quite reasonable up to \( p_{t} \)= 2GeV.
The calculation of the spectra at higher transverse momenta at SPS
energies is beyond the limits of present day computers, because of the 
very small cross section of hard process.

\section{Conclusions}

Employing the recently developed NE{\large X}US \textbf{3} model
where the parameters have been mainly fixed to the pp data at \( E_{lab}= \)
158 GeV, we compare its predictions on average quantities, longitudinal 
and transverse spectra between \( \sqrt{s} \) = 5 GeV and 65 GeV 
with the existing data of \( pp, np \)  and \( \bar{p}p \) collisions. 

We find a very nice agreement with data for $4\pi$ multiplicities,
rapidity and $x_F$ distributions as well as for transverse momenta. This
suggests that the basic mechanism of particle production is well described
in the NE{\large X}US \textbf{3} model. Based on this observation we can use
this model to study pA and AA collisions which will be the subject of
a forthcoming publication.

\end{document}